\documentclass[11pt, preprint]{aastex}

\newcommand{\sgra}{Sgr A}
\newcommand{\sgrastar}{Sgr A$^\star$}
\newcommand{\hcop}{HCO$^+$}
\newcommand{\kms}{\ifmmode{{\rm ~km~s}^{-1}}\else{~km~s$^{-1}$}\fi}

\newcommand{\Mo}{~$M_{\odot}$}
\newbox\grsign \setbox\grsign=\hbox{$>$} \newdimen\grdimen 
\grdimen=\ht\grsign
\newbox\laxbox \newbox\gaxbox
\setbox\gaxbox=\hbox{\raise.5ex\hbox{$>$}\llap
     {\lower.5ex\hbox{$\sim$}}}\ht1=\grdimen\dp1=0pt
\setbox\laxbox=\hbox{\raise.5ex\hbox{$<$}\llap
     {\lower.5ex\hbox{$\sim$}}}\ht2=\grdimen\dp2=0pt

\slugcomment{submitted to the Astrophysical Journal}
\shorttitle{Dense Gas Surrounding Sgr A$^\star$}
\shortauthors{Shukla, Yun, \& Scoville}

\begin{document}

\title{Dense Ionized and Neutral Gas Surrounding Sgr A$^\star$} 

\author{Hemant Shukla\altaffilmark{1}}
\affil{Owens Valley Radio Observatory, California
Institute of Technology, Pasadena, CA 91125} 
\email{hemant@ovro.caltech.edu}
 
\author{Min S. Yun}
\affil{University of Massachusetts, Astronomy Department, Amherst, 
MA 01003}
\email{myun@astro.umass.edu}
 
\author{N. Z. Scoville}
\affil{Owens Valley Radio Observatory, California
Institute of Technology, Pasadena, CA 91125}
\email{nzs@astro.caltech.edu}
 
\altaffiltext{1}{Present Address:  Stanford Research Systems, 
1290-C Reamwood Dr., Sunnyvale, CA, 94089; hshukla@thinksrs.com}
 
\medskip

\begin{abstract}
We present high resolution H41$\alpha$ hydrogen recombination line  
observations of the $1.2\arcmin$ (3 pc) region surrounding
Sagittarius A$^\star$ (Sgr A$^\star$) at 92 GHz using the 
Millimeter Array at the Owens Valley Radio Observatory (OVRO) with an 
angular resolution of $7\arcsec \times 3\arcsec$ and velocity resolution 
of 13 \kms.  New observations of H31$\alpha$, H35$\alpha$, 
H41$\alpha$ and  H44$\alpha$ lines were obtained using the NRAO 12-m 
telescope, and their relative line strengths 
are interpreted in terms of various possible emission mechanisms. 
These NRAO 12-m measurements are the most extensive recombination
line survey of this region to date. These OVRO data also represent
the highest angular resolution and the highest
sensitivity observations of \sgra\ West in hydrogen recombination
line and continuum emission at the millimeter (mm) wavelengths.
Observations of HCO$^+ (J = 1 \rightarrow 0)$ transition at 89 GHz
are obtained simultaneously with a 40\% improved angular resolution
and 4-15 times improved sensitivity over the previously published
results, and the distribution and kinematics
of the dense molecular gas in the circumnuclear disk (CND)
are mapped and compared with those of the ionized gas.  
The line brightness ratios of the hydrogen 
recombination lines are consistent with
purely spontaneous emission from $T_e \sim 7000$ K gas with 
$n_e \sim 2\times 10^4$ cm$^{-3}$ near LTE condition.  
A virial analysis suggests that the most prominent
molecular gas clumps in the CND have mean densities of order $10^7$
cm$^{-3}$, sufficient to withstand the tidal shear in the Galactic 
Center region.  Therefore, these
clumps may survive over several dynamical times, and the CND
may be a dynamically stable structure.  We estimate
a total gas mass of about $3\times 10^5 M_\odot$ for the CND.  
Our combined analysis of the new 
high resolution H41$\alpha$ and \hcop\ images and our kinematic model 
demonstrates a widely spread physical and dynamical 
link between the molecular gas in the CND and the ionized gas,
including along the eastern rim of the CND where a gap was
previously suggested.  

\end{abstract}


\section{Introduction \label{sec:intro}}

Studying radio recombination lines (RRLs) is one of the 
most direct ways of probing ionized gas such as HII regions. 
Ionized gas properties such as kinematics, electron temperatures, 
and the geometrical filling factor can be directly constrained 
by the recombination line
flux density, line width, free-free continuum flux density, and 
size of the line-emitting regions.  
At centimeter (cm) wavelengths, RRLs are primarily stimulated emission
amplifying extended non-thermal synchrotron emission.  Therefore, 
distinguishing emission arising from diffuse, low density
gas from that from clumpy, high density gas is difficult.  Also, pressure 
broadening of the line width can complicate the interpretation of the 
cm RRL observations. 
For instance, cm RRL emission from luminous starburst systems
such as Arp~220 may be dominated by internally 
stimulated emission from diffuse ionized gas \citep{Anantha93,Zhao96}
while spontaneous emission process is important at higher 
frequencies \citep[e.g. ][]{Anantha00,Yun03}. 
Interpretation of recombination lines at millimeter (mm) wavelengths
is more straightforward since stimulated
emission and pressure broadening become negligible in most cases. 
Recent studies of mm RRLs in Galactic H II regions 
\citep{Gordon90,Martin94,Gordon94} and starburst galaxies 
\citep{Seaquist96,Puxley97} found higher line fluxes than 
expected from the lower frequency measurements. 
A steep rise in line flux density with frequency, 
characteristic of optically thin spontaneous emission,
is seen in nearby starburst galaxies M82 and NGC~253 
\citep{Seaquist96,Puxley97}.

The central 1 parsec region surrounding the putative central
source in our Galactic Center, \sgrastar, is a frequent target
of multi-wavelength investigations primarily because it is the nearest
and the best laboratory for studying the active
galactic nucleus (AGN) phenomenon and its fueling process.
Currently this region appears to be relatively gas 
free. However, immediately surrounding \sgrastar\ is an 
array of ionized gas streamers \citep[``Sgr A West''; 
see review by][]{yusef00}.  These streamers have been previously 
studied in the 12.8 $\mu$m  [Ne II] \citep{Lacy80,Serabyn85},  
P$\alpha$ \citep{Scoville03}, and in
cm RRL \citep{Schwarz89,Roberts91,Roberts93,Roberts96}, and they 
consist of several distinct kinematic features. 
The total mass of the ionized streamers is estimated 
to be less than 100 M$_\odot$ in total.  Recent IR and radio 
observations found evidence for {\it neutral} gas within this central
``cavity'' \citep{Genzel85,Jackson93,Latvakoski99,Herrnstein02},
suggesting a more substantial presence of gas in the
central parsec region.  Surrounding and closely interacting with 
the ionized streamers is the dense and clumpy molecular ring 
commonly referred to as the circumnuclear disk (CND). The CND is 
asymmetric about \sgrastar, indicating its possible origin 
related to a gravitational
capture of a passing cloud or energetic disruption of a stable ring 
\citep{Morris96}. This ring is believed to represent a circumnuclear 
accretion disk feeding the inner parsec although some have argued
that the CND is not a complete ring \citep[see ][]{Marshall95,Wright01}. 
The ionized streamers in \sgra\ West
are explained as cloudlets from the clumpy ring that after losing 
momentum have fallen toward the center and thereby have stretched 
and have been ionized by photo-dissociation within the inner parsec 
\citep[see ][]{Sanders98}.  Sensitive imaging studies in cm RRL 
\citep{Roberts91} suggest high temperature ($T_e = 5000 - 20,000$ K), 
with a possible radial temperature gradient 
increasing toward \sgrastar.  The large proper motions of the stars very 
close to \sgrastar\ suggest central dark mass of 2.5-4.1 $\times 10^6$ 
M$_\odot$ \citep{Genzel99,Ghez00,Ghez03}.

Motivated by the detection of extragalactic mm RRLs and recent new
observations of the Galactic Center region, we have imaged the  
H$41\alpha$ line at high angular resolution toward \sgrastar\ in order to 
study the nature of the ionized streamers and their relation to the dense
neutral gas in and around the CND.  At high frequencies, most of the RRL
and continuum emission comes from the densest ionized gas \citep
[see ][]{Brown78,Mohan01}.  One of the primary goals of this study
is to examine the existence of ionized gas denser than that traced
by cm-wavelength RRLs in this region.
The nature of the radiation mechanism is investigated by analyzing 
the new H31$\alpha$, H35$\alpha$, H41$\alpha$, and 
H44$\alpha$ line observations obtained at the 
NRAO 12-m telescope at Kitt Peak, Arizona.  We further examine
whether any dense {\it neutral} gas is directly associated with 
the ionized streamers using our sensitive, high resolution 
HCO$^+$ (1--0) images.  Our HCO$^+$ (1--0) data have superior 
sensitivity and angular resolution over the previously published 
observations of similar molecular species 
\citep{Gusten87,Marr93,Wright01}.  In addition, a detailed analysis 
of the spatial distribution and kinematics of the dense molecular gas 
and ionized gas in the CND is offered using a simple kinematical model.

\section{Observations}

\subsection{OVRO Millimeter Telescope Array}

The H41$\alpha$ and HCO$^+$ (1--0) observations were made using the low, 
equatorial and high resolution configurations of the millimeter 
telescope array at Owens Valley Radio Observatory. 
The observational parameters from the multi-configuration 
observations conducted during the spring and fall 1998 are 
given in Table~\ref{tab:ovrosummary}. The OVRO array consists of 
six 10.4 meter telescopes equipped with cooled SIS receivers. 
Interferometric measurements are made using a total of 45
independent baselines in order to maximize the sampling of the
uv-plane and to achieve the highest imaging dynamic range possible.
In the high resolution configuration the longest baseline used was 220 
meters. The spectrometer modules were arranged so that each line was 
covered with a total of 60 channels at 4 MHz (13 \kms) resolution
centered at the velocity of $V_{LSR}=0.0$ \kms. 
By setting the LO frequency at 90.5 GHz, H41$\alpha$ line at 92 GHz 
(upper sideband) and HCO$^+$ (1--0) line at 89 GHz (lower sideband) 
could be observed simultaneously.  The primary beam (field of view) 
is about $85\arcsec$ at these frequencies. 
A nearby calibrator NRAO 530 was observed every 15 minutes to track 
the atmosphere and instrumental gain. The amplitude scaling is derived from 
the observations of 3C~273, whose flux density is monitored  
daily relative to Uranus and Neptune with an accuracy of about 20\%.

The data were reduced using the OVRO data reduction program 
MMA \citep{Scoville93}. \sgrastar\ is a low declination source, 
$\delta \sim -29^\circ$, and the synthesized beam is highly 
elongated in the N-S direction. The resulting beam size is 
$6.95\arcsec \times 3.47\arcsec$ (FWHM).
The data were imaged and analyzed using DIFMAP \citep{Pearson94} and 
Astronomical Image Processing System (AIPS) provided by the 
National Radio Astronomy Observatory (NRAO). 
Of the total 60 channels in the spectra, 15 line-free channels 
(60 MHz total) were used to construct the continuum image. Unresolved 
strong continuum source \sgrastar\ ($2.0\pm0.3$ Jy beam$^{-1}$) 
and the ionized 
streamers were used for self-calibration in DIFMAP, and the 
continuum was subtracted from the spectral line channels. 
All spectral and continuum images were corrected for primary 
beam attenuation. Subsequent analysis of the spectral line data 
cubes were conducted using
AIPS and Groningen Image Processing System (GIPSY).  

\subsection{NRAO 12-m Telescope}

We have obtained the H31$\alpha$ (210.502 GHz), H35$\alpha$ 
(147.047 GHz), H41$\alpha$ (92.034 GHz), and H44$\alpha$ 
(74.645 GHz) hydrogen recombination line spectra of Sgr A West
using the NRAO 12-m telescope at Kitt Peak during May 1998.  
The NRAO 12-m telescope is equipped with single sideband
SIS receivers, and typical system temperature ranged between
250 and 400 K (SSB), despite the low elevation of the source.  
The beam size is inversely proportional to the frequency and is 
69$''$ at 92 GHz.  Both the
filter banks (500 MHz at 2 MHz resolution) and hybrid digital
spectrometer (600 MHz at 0.78 MHz resolution) were used to
record the data.  A wide range of OFF positions were examined,
mostly along the minima in CO and HCN surveys of the Galactic
Center region.  After an extensive investigation, it was determined
that the hydrogen recombination line emission is highly localized 
around \sgrastar, and the final spectra presented here were obtained with 
a simple azimuth offset of $\pm3'$.

From the observatory report on the telescope
efficiency measurements, we adopt the corrected main beam efficiency 
$\eta^*_m$ of 0.48, 0.85, 0.90, and 0.95 for the four observed
frequencies, respectively \citep{Mangum00}.  
These efficiencies are used to convert 
the observed brightness temperature $T^*_R$ to the main beam 
temperature $T_{mb}$ and then to flux density.  The 
telescope efficiency measurements and the subsequent corrections are 
inherently uncertain, and we adopt an absolute uncertainty of 20\% 
for the measured flux density.

\section{RESULTS}

\subsection{Continuum at 92 GHz \label{sec:continuum}}

\begin{figure}
\epsscale{0.8}
\plotone{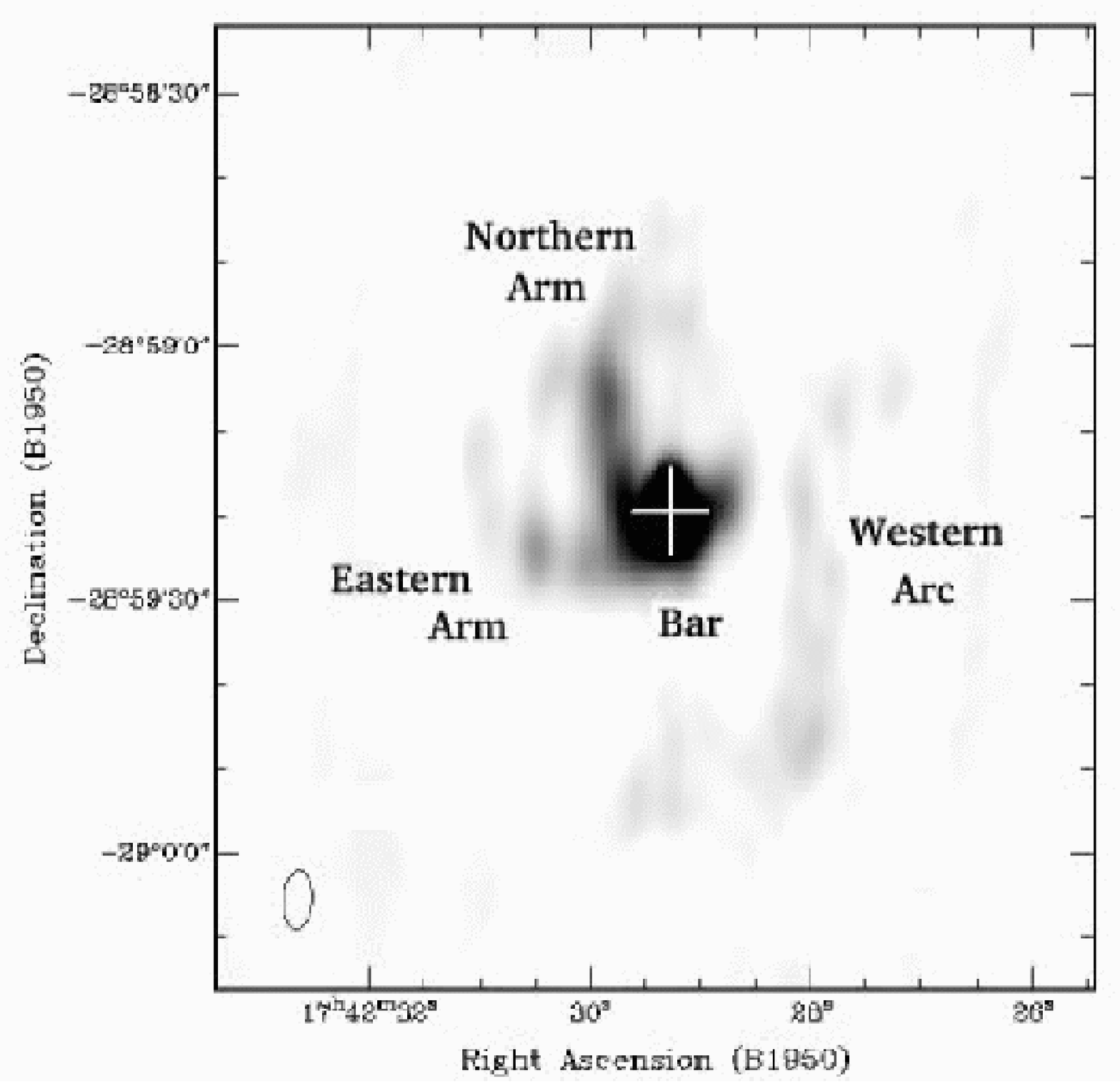}
\caption{Deconvolved image of 92 GHz continuum emission from \sgra\ complex at 
$6.95\arcsec \times 3.47\arcsec$ (PA = $-5^\circ$) resolution of the 
synthesized beam. The cross marks the position of \sgrastar. 
The grey-scale range is between 0.03 to 0.40 Jy beam$^{-1}$ 
with RMS noise of 0.01 Jy beam$^{-1}$.
\label{fig:continuum}}
\end{figure}

The 92 GHz continuum image of the \sgra\ complex at the 
resolution of $6.95\arcsec \times 3.47\arcsec$ (PA = $-5^\circ$)
is shown in Figure~\ref{fig:continuum}.  
The major components of the \sgra\ complex detected are 
the spiral structure known as Sgr A West and the compact central 
source \sgrastar, marked by a cross. The bright central source, 
\sgrastar, has peak flux density of $1.95\pm0.30$ Jy beam$^{-1}$. 
The total 92 GHz continuum flux density detected inside the 
85\arcsec\ diameter region is $\sim$12 Jy.  Some of the extended continuum
emission may be missed by our observations as the 19-pointing
mosaic image by \citet{Wright01} reported a total flux of
$\sim$15 Jy covering a slightly larger area while \citet{Wright93}
reported a total flux of 27 Jy within a 2\arcmin\ radius of \sgrastar\
from a combination of BIMA and Nobeyama 45-m telescope data.

The major continuum features of Sgr A West are identified in 
Figure~\ref{fig:continuum}, and their observed and derived
properties are tabulated in Table~\ref{tab:continuum}.
The east-west oriented filamentary structure 
located immediately south of \sgrastar\ is the bar of Sgr A West. The total 
continuum flux at 92 GHz from the bar is about 0.9 Jy with a peak 
brightness of 0.7 Jy beam$^{-1}$. 
The bar appears to connect to the western arc and the eastern arc 
on each ends. Running almost perpendicular to the bar is the northern 
arm with higher total integrated flux density of about 1.1 Jy. The 
continuum flux is higher toward the southern part of the northern arm. 
To the west of the bar is the western arc that curves south. It also extends 
north almost reaching the northern arm. The continuum flux from the western 
arc is relatively weak. The eastern end of the bar curves along the 
eastern arm but is not detected further north. The morphology of the 
ionized gas around the \sgra\ complex shown in 
Figure~\ref{fig:continuum} is remarkably 
similar to the 8.3 GHz continuum image by \citet{Roberts93}. 

The peak flux density measured outside \sgrastar\ range between 0.1 $-$ 0.75
Jy beam$^{-1}$ (see Table~\ref{tab:continuum}), which corresponds
to peak brightness temperature of 1.1 $-$ 6.0 K.  Assuming this continuum
is largely thermal free-free emission from 7000 K gas 
(see \S~\ref{sec:L/C}), one
can infer an optical depth of $\tau_c \lesssim 10^{-3}$ if the emitting
region fills the beam with a unit filling factor.  Even if the filling
factor is not near unity, this inference of
optically thin emission should be secure unless the continuum 
emission is entirely concentrated into discrete clumps with a
diameter of 0.005 pc (1000 AU) or smaller.  
Continuum optical depth is given by
\begin{equation}
\tau_c=0.0824~EM~T_e^{-1.35}\nu_{GHz}^{-2.1}.
\end{equation}
\noindent
Assuming $T_e=7000$ K, the emission measure 
($EM \equiv \int n_e^2 dl$) required to produce continuum opacity 
of $10^{-3}$ at 92 GHz is $2.5\times 10^7$ cm$^{-6}$ pc. If the 
thickness of the filaments along the line of sight is comparable to the 
apparent width ($l\sim 0.1$ pc), then the inferred mean
electron density is $n_e \sim 2\times 10^4$ cm$^{-3}$ 
($N_e\sim 5\times 10^{21}$ cm$^{-2}$).  If the ionized gas is 
clumped at much smaller scales, then the electron density
would be correspondingly larger.
In comparison, \citet{Lutz96} derived $n_e \sim 3000$ 
cm$^{-3}$ from their analysis of fine-structure line ratios.
The consideration of collisional de-excitation of the
observed [Ne II] emission places an upper limit on the electron
density of about 
$10^5$ cm$^{-3}$ \citep{Jackson93}.  Therefore the geometry
of the ionized gas must be fairly smooth over 0.01-0.1 pc scales.

\subsection {H41$\alpha$ \label{sec:h41a}}

The velocity integrated line flux density (``moment zero") image, shown in 
Figure~\ref{fig:H41a}, was generated by summing the continuum subtracted 
H41$\alpha$ image cube over all velocity channels with line emission 
(from $-293$ \kms\ to +221 \kms; see Figure~3). The cross 
marks the position of \sgrastar. 
The main features seen in the line emission map are very similar to the 
92 GHz continuum map shown in Figure~\ref{fig:continuum}, 
except for the absence of the 
bright, compact source corresponding to \sgrastar. The emission from the 
western arc is weaker than seen in continuum.
The northern arm is the brightest continuum feature besides 
\sgrastar, and the brightest H41$\alpha$ emission region also occurs 
in the northern arm just northeast of \sgrastar. The central bar shows 
a wide range of emission intensity and structures.

A summary of the measured parameters for all major features are 
given in Table~\ref{tab:h41a}. The sizes tabulated for the listed 
regions have been calculated 
assuming 1$\arcmin$ = 2.5 pc \citep{Reid89}. Unlike the 92 GHz continuum 
that peaks at \sgrastar, the brightest feature in H41$\alpha$ occurs in the 
northern arm with a peak brightness temperature of 
$\Delta T_{peak}=9.4\pm2.2$ K.  The western arc, however, 
appears relatively weak in H41$\alpha$ line. 
The peak line brightness for all H41$\alpha$ are $\le 10$ K,
and thus the H41$\alpha$ emission is also optically thin,
$\tau_L\lesssim 10^{-3}$ (see below).

The spatially averaged H41$\alpha$ line profiles for the six prominent 
emission regions (identified with boxes) are shown in Figure~\ref{fig:L/C}.  
The GIPSY tasks PROFIL and PROFIT were used to calculate the 
line-to-continuum (L/C) ratios and to fit a Gaussian profile to the data 
using a least squares procedure. The crosses indicate the data points, 
the solid lines are the Gaussian profiles, and the dashed lines 
represent the residuals.  Calculations of Gaussian profiles yield
four parameters: peak line-to-continuum flux density ratio, 
velocity integrated line intensity, central 
velocity, and the line width. These parameters are summarized in
Table~\ref{tab:Te} along with the derived LTE electron temperature
derived from the L/C ratios (see further discussions in \S~\ref{sec:L/C}). 
A typical L/C ratio along the northern arm is around 0.4. 
The L/C ratio is the largest ($\sim 0.9$) in the western arc while 
the L/C ratio in the bar drops from 0.6 to 0.03 within the
2 to 4 beam radii of \sgrastar.  

The kinematics of the ionized gas traced by H41$\alpha$ shown in 
Figures~3 and \ref{fig:L/C} are generally consistent with that of rotation 
about \sgrastar, similar to that of the CND, in broad
agreement with previous studies of SgrA West in RRLs \citep{Roberts91,
Roberts93} and in 12.8 $\mu$m [Ne II] \citep{Serabyn85}.  
The central bar spans a very wide range of velocity, between  
+33 \kms\ and $-$293 \kms\ (see Figure~3), much larger
than those associated with the northern arm (between +163 \kms\ 
and +7 \kms) and the western arc (between $-$7 \kms\ and $-$111 \kms).
This large velocity gradient suggests that this gas is located 
deeper in the gravitational potential, much closer to the central
mass concentration, than the rest of the ionized gas.  
Citing the relatively cool dust temperature along the bar, 
\citet{Latvakoski99} have argued that the nearness of the bar
to \sgrastar\ is largely in projection only, but this large spread
in velocity and the correspondingly large velocity gradient requires
that this gas feature has to be located much closer to the central
mass concentration than the northern arm and the western arc.  
Since the CND is already viewed with a large inclination 
($i \sim 60^\circ$), placing a gas streamer in a more edge-on orbit
can increase the line-of-sight velocity only by about 25\%.  Therefore
a viewing geometry alone cannot account for the large observed velocity
gradient.  Neither the angular resolution nor the
overall S/N of the new H41$\alpha$ data are better than the previous
VLA H92$\alpha$ measurements, and the main new insight 
obtained is that the ionized gas traced 
in H41$\alpha$ emission share the same kinematics as the gas
responsible for the H92$\alpha$ emission.

The H41$\alpha$ line widths of the ionized streamers range between 
43 \kms\ and 347 \kms\ (see Table~\ref{tab:Te}). 
The broadest line is seen toward \sgrastar\ with $\Delta$V
= $347\pm48$ \kms\ (see Figure~\ref{fig:L/C}, Box 4), but the S/N of this 
measurement is quite small in each individual channels. The next 
broadest line occurs in the vicinity of bright H41$\alpha$ 
region in the northern arm with $\Delta$V $= 213\pm 18$ \kms\ 
(Box 6 in Figure~\ref{fig:L/C}).  The narrowest line is seen
toward the southern end of the western arc (Box 3 in 
Figure~\ref{fig:L/C}) with $\Delta$V $= 43\pm5$ \kms.
Using 2\arcsec\ resolution VLA H92$\alpha$ images, \citet{Roberts93}
have shown that crowding of distinct kinematical components along
the line of sight can make the emission lines appear
much broader than the intrinsic line width of about 50 \kms.
Combined with the large velocity gradient expected near the
central mass concentration, the large line width we derive
from H41$\alpha$ emission must be primarily the result of spatial 
averaging resulting from our large observing beam of 
$7.0\arcsec \times 3.5\arcsec$.  A typical line width measured in
each pixel is between 40-75 \kms.
Given the low S/N of the data, it is not physically meaningful
to deconvolve the individual kinematic components by fitting 
multiple Gaussians to the observed spectra.  
There is a clear trend of increasing L/C ratio with a decreasing
line width.  Such a trend is expected if the line and continuum
emission arises from the same ionized gas clouds and the emitting 
gas has a narrow range of electron temperature (see \S~\ref{sec:L/C}).

\subsection {HCO$^+  (J = 1 \rightarrow 0)$ \label{sec:hco+}}

Figure~\ref{fig:hco+} shows the high resolution image of the molecular 
ring in HCO$^+$ emission (grey-scale) superimposed over the H41$\alpha$ 
(contour) emission. The circumnuclear ring traced in HCO$^+$ appears 
more complete than seen in the previous HCN ~\citep{Gusten87} and 
HCO$^+$ ~\citep{Marr92} images obtained using the BIMA array. 
More recently \citet{Wright01} imaged HCN, HCO$^+$, and H42$\alpha$ in a 
wider field of view using a 19-pointing mosaic, offering a much
better overall picture of the molecular gas distribution in the 10 pc
region surrounding the \sgrastar.  However, the latest BIMA data 
by Wright et al. are not much better than the older BIMA data in resolution 
and sensitivity.  These different images agree with each other 
broadly, but significant differences also exist in the details.  
One important difference is that the angular resolution of 
our new OVRO data is 40\% better than those of G\"{u}sten et al. and 
Marr et al., and our sensitivity is also 4 and 15 times improved, 
respectively.  The BIMA array is more sensitive to faint extended
emission, and its data reduction procedure is tuned to emphasize 
such features.  In contrast, our moment analysis is
more sensitive to identifying bright emission clumps,
and this accounts for the more complete appearance of the CND
in Figure~\ref{fig:hco+}, consisting of many discrete features.  

Determining whether any dense {\it neutral} gas is associated with the 
ionized streamers surrounding \sgrastar\ is one of the major objectives
for our study.  While the new observations have revealed clear
evidence for a close physical link between ionized and neutral molecular
gas (as discussed later in \S\ref{sec:CND}), no HCO$^+$ emission is found 
spatially and kinematically coincident with the ionized streamers 
traced in H41$\alpha$ (See Figure~3).
Actually HCO$^+$ absorption is seen in many channels,
particularly at blueshifted velocities, but most of these features
show little kinematic relation to the H41$\alpha$ emission.
The spatial distribution of HCO$^+$ absorption resembles the entire
continuum structure, and therefore the absorbing gas must be located
in the foreground, associated with the 100 pc scale molecular disk imaged 
by \citet{Tsuboi99} and others.  
Among the redshifted velocity channels where little
foreground absorption is present (e.g., +117 to +195 km/s), there
is little evidence for HCO$^+$ absorption at the locations of 
the H41$\alpha$ and continuum emission peaks.  
In these channels, the upper limit to the optical depth for HCO$^+$ 
absorption is about 0.3 ($3\sigma$)
at a spectral resolution of 13 \kms.  This limit corresponds to
an upper limit on the HCO$^+$ column density of $4\times 10^{12}$ 
cm$^{-2}$.  For an assumed HCO$^+$ abundance of $10^{-9}$, this
translates into an H$_2$ column density of $4\times 10^{21}$ 
cm$^{-2}$. If the HCO$^+$ abundance appropriate for a translucent 
cloud \citep[where ion chemistry is important, see ][]{hoger95} is adopted, 
the inferred upper limit on the $H_2$ column density becomes as
little as $\sim 10^{19}$ cm$^{-2}$.  These upper limits are close
to or smaller than the column density of $N_H\sim (0.5-1.0) \times 
10^{22}$ cm$^{-2}$ inferred from the [O I]/[C II] far-IR line
ratio \citep{Jackson93} and the mid-IR color analysis 
\citep{Latvakoski99}.  Using the calculations by \citet{Sternberg89},
\citet{Jackson93} have proposed that most of the molecules in
the neutral gas inside the central cavity with $N_H \lesssim 10^{22}$ 
cm$^{-2}$ are photo-dissociated and that the neutral gas is primarily atomic.  
Our HCO$^+$ absorption measurements are consistent with
these conclusions, and we conclude independently that Sgr A West 
filaments are either highly ionized or are largely atomic.

\subsection{H31$\alpha$, H35$\alpha$, H41$\alpha$, and H44$\alpha$
\label{sec:multi-lines}}

The hydrogen recombination line spectra obtained at the NRAO 12-m
telescope are shown in Figure~\ref{fig:12m}, and the observing frequency 
and the measured line flux are summarized in Table~\ref{tab:12m}.  
The mm frequency window is rich in spectral transitions, and 
studying faint emission lines in the Galactic Center region 
can be challenging. The H35$\alpha$ line is not
shown because it is completely swamped by the 
much stronger CS $J=3\rightarrow 2$ line at 146.969 GHz, and no useful 
information could be derived.

The H44$\alpha$ spectrum, shown at the bottom of Figure~\ref{fig:12m}, 
is the least confused and the highest S/N measurement among
all transitions observed at the 12-m telescope.  The observed line
profile closely resembles the H41$\alpha$ spectrum derived from
the OVRO interferometer data (shown with a solid line in the
middle of Figure~\ref{fig:12m}).  
The total velocity integrated line flux density
is $232\pm46$ Jy km s$^{-1}$ within the 84\arcsec\ beam of the
12-m telescope at 74.645 GHz.

The H41$\alpha$ spectrum obtained using the NRAO 12m telescope,
shown with a dotted line in the middle of Figure~\ref{fig:12m},
shows a bright, narrow feature near +200 km s$^{-1}$.  This is
not seen in the H41$\alpha$ spectrum obtained from the OVRO data
(shown in solid line).  We have mapped H41$\alpha$ 
emission in a 2\arcmin\ by 3\arcmin\ region centered on \sgrastar\ in
order to examine the spatial extent of this spectral feature.
The broad emission feature seen in the OVRO data and
in the H44$\alpha$ spectrum is spatially confined only to the central
position while the bright, narrow feature appears everywhere
with little velocity gradient.  The narrow line width, 
little velocity gradient, and wide-spread distribution  
strongly indicate that this narrow component
is a Galactic foreground emission.  
A likely identification is CH$_3$CN, which has transitions at 
91.959, 91.970, 91.980, 91.985, and 91.987 GHz.
The total velocity integrated H41$\alpha$ line flux density from the 
OVRO data is $152\pm30$ Jy \kms.  This is a lower limit since some 
extended flux may be missed by the interferometer (see below).

The H31$\alpha$ spectrum shown at the top of Figure~\ref{fig:12m} is obtained 
by summing 8 independent spectra covering the whole Sgr A West complex.
The telescope beam at 210.502 GHz is about 30\arcsec\ and spatially
resolves the Sgr A West structure.  Therefore a 3 $\times$ 3 grid 
map was obtained at full beam sampling.  The NE grid point
was not observed, and the 8 spectra are summed together
to derive the integrated H31$\alpha$ spectrum.  The H41$\alpha$ channel maps
(Figure~3) suggest that our H31$\alpha$ spectrum may
be missing some redshifted emission at $V$ = 50 - 150 km s$^{-1}$.
Nevertheless, the summed spectrum closely resembles the H41$\alpha$ and 
H44$\alpha$ spectra, and we made no attempt to correct for any missing
flux.
 
\section{DISCUSSION}

\subsection {Massive Compact Molecular Clumps in the CND 
\label{sec:clumps}} 

The nine brightest HCO$^+$ clumps within the CND are identified 
as localized peaks of velocity integrated HCO$^+$ flux density map 
(Figure~\ref{fig:hco+}), and their physical sizes and 
emission characteristics are summarized in Table~\ref{tab:hco+}. The peak 
line brightness temperature, averaged over the $7''\times 3.5''$ 
(0.29 pc $\times$ 0.15 pc) beam, ranges between 10 and 30 K.  
In comparison, the gas temperature derived from NH$_3$ is 60-70 K 
\citep{Coil00} while the dust temperature of about 75 K is
derived by Latvakoski et al. (1999) from the far-infrared observations.
Since HCO$^+$ emission is strongly absorbed by the foreground gas along the 
line of sight and the beam filling factor is likely less than 
unity, the gas temperature inferred from the observed line
brightness is strictly a lower limit.  The measured peak HCO$^+$ line 
temperature is comparable to the values found by earlier BIMA studies 
(see \S~\ref{sec:hco+}) with a lower angular resolution, and one can 
safely deduce that the optically thick molecular gas nearly fills the 
beam at $\sim 0.1$ pc scales in these CND clumps.

We have computed virial masses of the individual HCO$^+$ emitting
clouds as,
\begin{equation}
M_{vir} = 250~r \Delta V^2~M_\odot ,
\end{equation}
where $r$ is the radius of the cloud in pc and $\Delta V$ is the line width
in km s$^{-1}$.  If these individual 
molecular gas clumps are self-gravitating, then their masses range between 
$(0.3-4.5)\times 10^4 M_\odot$ (see Table~\ref{tab:hco+}).  
Assuming a spherical geometry, the mean mass density ($\bar{\rho} 
\equiv M_{vir}/ \frac{4}{3}\pi r^3$) is between 
$(1.3-10)\times 10^{-17}$ g cm$^{-3}$ and mean H$_2$ molecule density of
$\bar{n}_{vir} = (0.4-3.0)\times 10^7$ cm$^{-3}$.  
These densities are significantly
larger than the values inferred from the excitation analysis 
\citep[$n \sim 10^{5-6}$ cm$^{-3}$,][]{Genzel85,Marr93,Marshall95}.  
For a central source with a mass of 
$4.1\pm0.6 \times 10^6 M_\odot$ \citep{Ghez03}, 
the critical mass density required to survive the tidal shear 
at a radius of 1.5 pc is $\bar{\rho}_{tidal}=4.3\times 10^{-17}$ 
g cm$^{-3}$ or $\bar{n}_{tidal}=2.5\times 10^7$ cm$^{-3}$.  
Contrary to the previous studies using excitation analysis, we 
conclude that the mean gas density inferred from the virial 
analysis is close to or exceeding the critical density for surviving 
the tidal shear if they are self-gravitating.
Also, the CND may well be a structure that can survive several
dynamical time scales ($\gtrsim 10^5$ years). 
The total gas mass for the CND inferred from our virial analysis
is around $3\times 10^5 M_\odot$, which is only a few per cent
of the total dynamical mass for the central one parsec region.
The virial mass of each of these clumps are about two orders of
magnitude larger than that of the ionized gas associated with 
the whole ionized spiral (Sgr A West).  Therefore, an infall of 
even a much smaller gas clump is sufficient to account for the observed 
ionized streamers. The new higher resolution 10-field mosaic 
observations of HCO$^+$ and HCN emission by \citet{Christopher03}
suggest even higher mean density ($\bar{n}\sim 10^8$ cm$^{-3}$) 
for these dense neutral clumps.

\subsection {Electron Temperature in Sgr A West
\label{sec:L/C}} 

\begin{figure}
\plotone{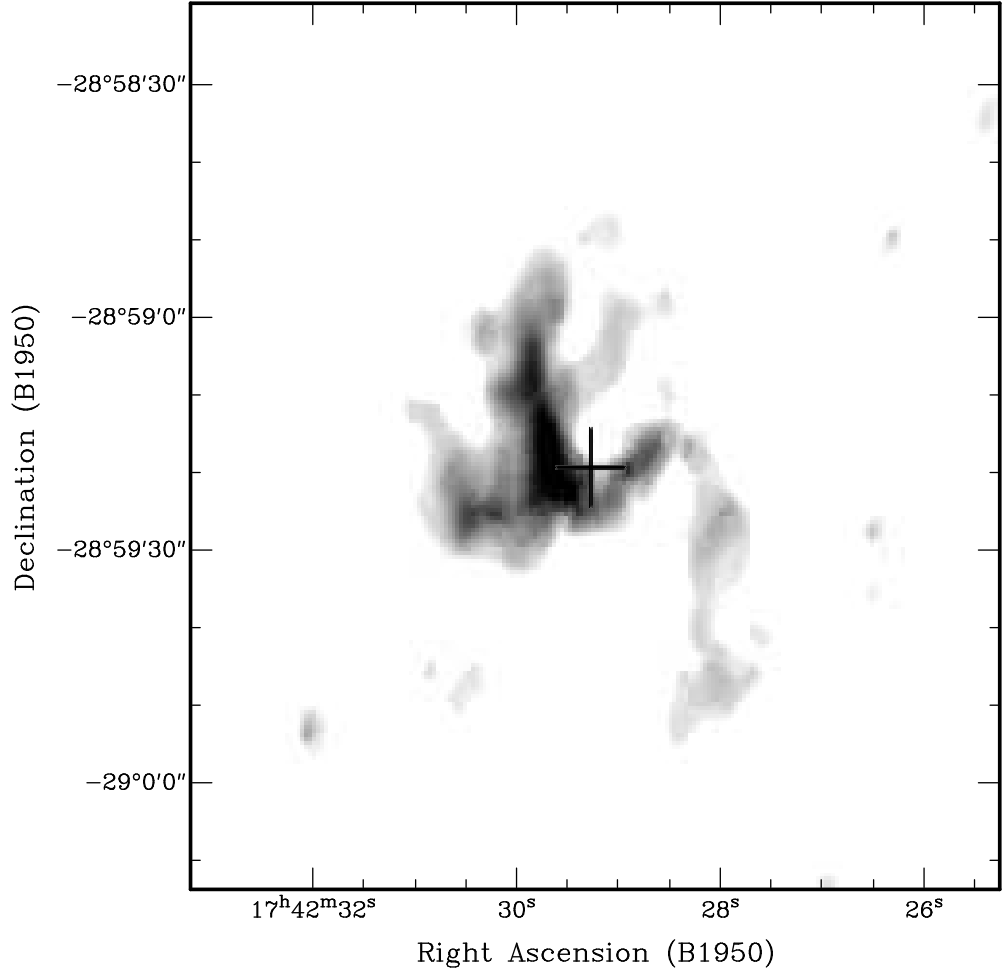}
\caption{Integrated line emission of the H 41$\alpha$ line from 
\sgra\ complex at 
$6.95\arcsec \times 3.47\arcsec$ (PA = $-5^\circ$) resolution of the 
synthesized beam. The cross marks the position of \sgrastar. 
The grey-scale range is between 0 and 12 Jy km s$^{-1}$ beam$^{-1}$.
\label{fig:H41a}}
\end{figure}

The 92 GHz continuum around \sgra\ is primarily free-free emission, and 
a good spatial correspondence between the continuum 
(Fig.~\ref{fig:continuum}) and the H41$\alpha$ line (Fig.~\ref{fig:H41a})
emission is naturally expected. Despite the similar 
morphology, however, a comparison suggests that the line 
emission is significantly clumpier than the continuum, especially 
in the western arc. The central bar is also clumpier in the line 
emission.  In addition, in the northern arm the line emission appears 
brighter than anywhere else with an increasing trend toward \sgrastar. 
These differences may not be intrinsic to the source.
A lower signal-to-noise ratio in the line image
probably accounts for some of the differences.  In addition,
the continuum image suffers more from the missing flux problem than 
the line images because of a larger effective source 
size.\footnote{A line emitting area is significantly smaller in size 
than the continuum emitting region in individual channel maps for an 
object with a large velocity gradient such as a rotating disk.}

For optically thin gas in LTE emitting thermal continuum and negligible 
pressure broadening, LTE electron temperature $T^*_e$ can 
be derived from the line-to-continuum ratio as, 
\begin{equation}
T^*_e = (6943 \,\nu^{1.1} \,{S_C\over {S_L \Delta V}} \,\, 
{1 \over {1 + Y^+}})^{0.87} 
\end{equation}
where $\nu$ is the observing frequency in GHz, $\Delta V$ is the velocity 
width  (FWHM) in km s$^{-1}$, $S_C$ and $S_L$ are continuum and line flux
densities, respectively, and $Y^+$ is the singly ionized helium 
abundance \citep{Brown78,Roberts93}. 
In the case of Sgr A West the aforementioned assumptions may be 
reasonably made along with the interpretation that the continuum at 
92 GHz is largely free-free emission. The singly ionized helium  
abundance $Y^+$ of 10\% is assumed for Sgr A West \citep{Roberts93}. 
The H41$\alpha$ line widths listed in Table~\ref{tab:Te}
are larger than the typical H92$\alpha$ line widths of 40-50 \kms\
reported by \citet[][]{Roberts93}, and this is the direct
consequence of a poorer angular resolution of the H41$\alpha$ 
observations.  The resulting blending of emission features could conspire
to produce unusual L/C ratios if gas temperature and density vary
significantly along the ionized streamers.  The high angular 
resolution observations by \citet[][]{Roberts93} have 
shown that this is not the case, and the poorer
angular resolution of our H41$\alpha$ observations should not
affect our $T_e$ analysis in most cases.

An electron temperature map can be derived from the 92 GHz continuum
map (Fig.~\ref{fig:continuum}), the velocity integrated H41$\alpha$
map (Fig.~\ref{fig:H41a}), and the relation given in Eq.~3.
The derived electron temperature map, shown in Figure~\ref{fig:Te}, 
reveals a fairly uniform distribution of $T_e$ around 7000 K 
throughout the Sgr A West complex 
except near \sgrastar\ (see Table~\ref{tab:Te}).  
In general $T_e^*$ derived at higher frequencies 
tends to be larger because of diminishing contribution from  
stimulated emission.  However, electron temperature derived
with H41$\alpha$ line at 92 GHz agrees well with the 
derivation using H92$\alpha$ line at 8.3 GHz 
\citep[$7000\pm500$ K, ][]{Roberts93}, suggesting that the
contribution from stimulated emission is negligible even 
frequencies as low as 8.3 GHz.  
Detecting little H41$\alpha$ emission in the direction of the 
bright continuum source \sgrastar\ is another clear
evidence that stimulated emission is not important at 92 GHz
(see \S~\ref{sec:RRL}).

The higher electron temperature derived along the central bar
just south of \sgrastar\ has previously been noted by \citet{Schwarz89} 
and \citet{Roberts91}, and whether this is a real effect or an artifact 
of data analysis has been a subject of recent debates.  Citing the 
discovery of highly blueshifted line emission near $-250$ \kms\ missed 
by these earlier studies, \citet{Roberts93} have suggested that
the unusually low L/C ratio and thus
the inferences of higher electron temperature may be erroneous. 
In general, a lower than expected L/C ratio can result from either
(1) missing line flux or (2) extra continuum flux.  The explanation
by \citet{Roberts93} is the former, but a careful analysis
using the actual measurements of the extremely blueshifted line
by \citet{Roberts96} revealed that the missed flux alone cannot
fully account for the low L/C ratio.  In addition, \citet{Roberts96} 
noted the presence of two optically thick continuum sources IRS~2 and
IRS~13 whose additional continuum flux should contribute to the 
unusually low L/C ratio (and higher $T_e$) observed.
A broad coincidence of this position with that of the ``mini-cavity" 
\citep{Yusef90} and its possible impact on the calculation of 
$T_e^*$ are discussed in detail by \citet{Roberts93}.  
Presence of significant structures 
at scales smaller than our synthesized beam violates the 
simple geometrical assumption going into Eq.~3 \citep[also see][] 
{Brown78}, and this may account for part of 
the observed local ``enhancement'' in $T_e^*$.  

The local enhancement in electron temperature 
near the location of IRS~2 and IRS~13 is clearly seen 
in Figure~\ref{fig:Te} in the same manner as the seen in Figure~14 of
\citet{Roberts93}.  The persistence of this feature at
92 GHz implies that the 
presence of the two continuum sources as suggested by 
\citet{Roberts96} does not offer a full explanation for the
observed low L/C ratio.
\citet{Roberts96} estimated an optical depth of $\sim 0.8$ and
$\sim 1.1$ at 8.3 GHz for the continuum sources 
IRS~2 and IRS~13 by comparing their 8.3 GHz image with the
5 and 15 GHz VLA continuum images.   Free-free optical depth scales 
as $\tau \propto \nu^{-2.1}$ (see Eq.~1), and the continuum optical
depth for IRS~2 and IRS~13 should be $\lesssim 0.01$ at 92 GHz.
In comparison, optically thin free-free emission associated
with the ionized filament should scale only as $S_\nu \propto \nu^{-0.1}$.
As a result, the continuum contribution from IRS~2 and IRS~13 should be 
substantially reduced at 92 GHz, and the L/C ratio should reflect mostly 
the optically thin emission associated with the ionized gas
filaments in SgrA West.  However, the observed L/C ratio at 92 GHz is 
still noticeably lower at this location.
The area of a lower L/C ratio is not centered on these 
two continuum sources either.  If continuum opacities for these two
sources are much larger than estimated by Roberts et al. (i.e., 
optically thick even at 92 GHz),
then the explanation of IRS~2 and IRS~13 as additional sources of 
continuum is still plausible, although the spatial distribution is
still a problem. 
In summary, the apparent enhancement in $T_e$ (a low 
L/C ratio) along the central bar just southwest of \sgrastar\
is still not fully explained.

\subsection {Nature of the Recombination Line Emission \label{sec:RRL}}

The flux density ratio between H41$\alpha$ and H92$\alpha$ lines
is shown in Figure~\ref{fig:H92a/H41a}.  The overall structure of
the northern arm, the bar, western arc, and eastern arm is almost
identical at both frequencies, and the line ratio is nearly 
constant, $S_{H92\alpha}/S_{H41\alpha} \sim0.15$.
This similarity is remarkable given the order of magnitude 
difference in the frequencies of these lines. 
A slightly higher ratio is seen along the western arc and
toward the top where the western arc arches into the northern 
arm, suggesting brighter emission at centimeter wavelengths. This 
trend of comparatively brighter emission at longer wavelength 
was also noted in the comparison of H92$\alpha$ to 
Br$\gamma$ ratios by \cite{Roberts93}. 
Roberts \& Goss offered extinction of Br$\gamma$ emission
by dust in the CND as a possible explanation, but a higher
$S_{H92\alpha}/S_{H41\alpha}$ ratio would require a different
explanation.  Since the derived $T_e^*$ distribution is fairly constant 
(see Fig.~\ref{fig:Te} and Table~\ref{tab:Te}), the enhanced
$S_{H92\alpha}/S_{H41\alpha}$ ratio is not likely caused by
any changes in the physical characteristics of the gas.  
The comparison of the total integrated H41$\alpha$ line flux with 
those of other transitions measured with the NRAO 12-m telescope 
suggest that some extended flux is missed by the 
interferometer (see \S~\ref{sec:multi-lines}).  
The shortest baseline in the OVRO 
observations of 3100 $\lambda$ is about 4 times larger than that 
for the VLA observations (830 $\lambda$), and the greater missing 
flux in the OVRO data offers a plausible 
explanation for the apparent enhanced $S_{H92\alpha}/S_{H41\alpha}$ 
ratio along the western arc.

The analysis of the 92 GHz continuum (\S\ref{sec:continuum}) and
the H41$\alpha$ emission (\S\ref{sec:h41a}) suggests high electron 
density ($n_e \sim 2\times 10^4$ cm$^{-3}$) and low optical depth 
($\tau \lesssim 10^{-3}$) for the ionized streamers.  Under these 
conditions, both internal and external stimulated emission should be
negligible, and optically thin spontaneous emission should dominate 
the recombination line emission at millimeter wavelengths. 
For the optically thin case near LTE, Eq.~3 implies that 
the velocity integrated line flux density $S_L \Delta V$ should 
increase linearly with frequency since $S_C \propto \nu^{-0.1}$ 
for thermal free-free emission.\footnote{Using slightly different 
assumptions, \citet{Seaquist96} derive a relation showing 
$S_L \Delta V \propto \nu^{+2}
$ in their Eq.~A9.  However, the 
$n^3$ factor, which they treated as a ``constant" term going 
from their Eq.~A8 to Eq.~A9, actually behaves like $\nu^{-1}$
($\nu_{Hn\alpha} \propto [1/n^2 - 1/(n+1)^2]$), and this too leads 
to a linear relation, $S_L \Delta V \propto \nu^{+1}$.}
A plot of the velocity integrated line flux density for 
H31$\alpha$, H41$\alpha$, H44$\alpha$, and H92$\alpha$ lines is
shown as a function of frequency shown in Figure~\ref{fig:LTE},
and indeed a nearly linear trend is seen.  
Excluding the H41$\alpha$ measurement from the OVRO (which 
is a lower limit because of the missing flux), a best fit 
power-law slope is $\alpha=+0.95 \pm 0.18$, which is nearly
exactly the theoretically expected value.  

The assumption of ``near LTE'' is not strictly correct, however. A general 
expression for recombination line brightness in comparison
to the LTE value is
\begin{equation}
T_L = T_L^* b_n (1-\frac{1}{2}\tau_c \beta_n)
\end{equation}
\noindent
where $T_L^*$ is the LTE line brightness, $b_n$ is the departure
coefficient, $\tau_c$ is continuum optical depth, and $\beta_n$
is the effective line absorption coefficient accounting for
non-LTE stimulated emission in the line.  For the range of
physical conditions we derive for the \sgra\ West ($T_e \sim 7000$ K,
$n_e \sim 10^4$ cm$^{-3}$), the computed values of these parameters 
are $b_n \sim$ 0.7-0.9 and
$(1-\beta_n) \sim 25$ \citep{Brocklehurst70,Walmsley90}.
Although the non-LTE factor in the line absorption coefficient
$\beta_n$ is large, it has little impact on the observed line
brightness because a small optical depth ($\tau_c \lesssim 10^{-3}$,
as shown in \S~\ref{sec:continuum}) makes the second term in Eq.~4
negligible, i.e. $\frac{1}{2}\tau_c \beta_n \sim 0$.  The net effect
is that the observed recombination line brightness is reduced by a factor
of $b_n$ from the LTE value (i.e., $T_L/T_L^* = b_n$).
The linear slope of the the observed power-law relation between the
velocity integrated line flux density and frequency is unaffected
as long as $\frac{1}{2}\tau_c \beta_n \sim 0$.  As a result, the average
flux density of the recombination line should increase linearly with 
frequency even when some non-LTE effects are present as long as  
spontaneous emission dominates the radiation process.

The comparison of the H41$\alpha$ and H92$\alpha$ spectra obtained
toward Sgr A$^\star$ (Figure~\ref{fig:twospecs}) offers a
definitive proof that the narrow H92$\alpha$ line seen near 
$V=+50$ \kms\ by \citet{Roberts93} is indeed stimulated emission. 
Citing the nonthermal nature of the Sgr A$^\star$ continuum emission
and the low observed L/C ratio, Roberts \& Goss suggested
that this H92$\alpha$ emission must be stimulated emission
arising from diffuse foreground ionized gas.  
The H92$\alpha$ spectrum (shown in
dotted line) and the H41$\alpha$ spectrum (shown in solid line) 
have essentially identical shape at a position 
along the Northern Arm, 7\arcsec\ offset to the east of Sgr A$^\star$,
suggesting the same ionized gas is responsible for both 
spectral features.  At the location of Sgr A$^\star$, however,
H92$\alpha$ spectrum shows a distinct narrow component near $V=+50$ 
km s$^{-1}$, not seen in the H41$\alpha$ spectrum.  
As shown in Eq.~1, the optical depth for H41$\alpha$ line is
expected to be more than 100 times smaller than the H92$\alpha$ line, and
H41$\alpha$ is expected to be smaller by this large factor
for optically thin stimulated emission.

\subsection{Physical Link between Neutral and Ionized Gas 
\label{sec:CND}}

A review of the recent literature yields a coherent but complex
picture for the composition, geometry, and kinematics of the neutral
and ionized gas found in the central few parsec of our Galaxy.
The CND, traced in \hcop\ and HCN, has a ring-like morphology 
with an inner radius of about 30\arcsec\ (1.25 pc) and an outer
radius of at least 45-60\arcsec\ (1.8-2.5 pc) and 
is physically and kinematically linked with a larger scale disk with
an extent of 10 pc or more \citep{Wright01,McGary01,Christopher03}.
It may be made of a multi-phase medium as the estimates of gas
temperature (50-250 K) and density ($10^{4-6}$ cm$^{-3}$ and
$ \gtrsim 10^7$ cm$^{-3}$ in dense clumps, see \S~\ref{sec:hco+})
range widely \citep{Marr93,Jackson93,Marshall95,Latvakoski99}.
Within it lies the ``central cavity'' which actually contains
the ionized streamers Sgr A West, made of about 100 \Mo\ of
$n_e\sim 10^4$ cm$^{-3}$ gas at $T_e\sim 7000$ K and as much
as 1000 \Mo\ of neutral gas which is thought to be mostly
atomic with $n_H \sim 10^5$ cm$^{-3}$ and T = 100-300 K
\citep[][ and see \S~\ref{sec:hco+}]
{Genzel85,Roberts93,Jackson93,Telesco96,Latvakoski99}.
By obtaining the most sensitive and the highest angular resolution 
images of H41$\alpha$, 92 GHz continuum, and \hcop\ (1--0) 
to date simultaneously, we are able to shed some new light on the 
physical link between the neutral and ionized gas in this region.

The clearest demonstration of the physical connection between 
the neutral and ionized gas is the presence of recombination 
line peaks along the western arc, 
just inside of the dense gas clumps within the CND as shown
by the overlay of the velocity integrated H41$\alpha$ 
and HCO$^+$ images (see Figure~\ref{fig:hco+}).
This radial offset between the ionized streamers 
and the molecular ring has been noted previously \citep{Roberts93},
and it is consistent with the suggestion that the western arc is
the ionized inner edge of the molecular ring directly exposed 
to a central ionizing source \citep[][]{Telesco96}.  
Evidence for a similar link between neutral and ionized gas
is also seen on the eastern rim of the CND, between the 
eastern arm and the clump I shown in Figure~\ref{fig:hco+}.  
From their comparison of the 8 GHz radio continuum map and 
the 30 $\mu$m optical depth map, 
\citet{Telesco96} postulated that the eastern arm is the ionized 
front of a protrusion of high-density material facing the central 
ionizing source.  The clump I identified in Figure~\ref{fig:hco+}
may indeed be this previously unseen high-density material, 
revealed for the first time.  The extinction map derived from
the HST/NICMOS P$\alpha$ observations by \citet{Scoville03}
shows a hole in the obscuring material towards
the northern arm and western arm, but high extinction regions with
$A_V\ge30$ surround this hole, including the HCN clump
corresponding to our clump I.  Possible ionizing source
candidates range from the putative central engine \sgrastar\ to hot UV 
stars near the Galactic center \citep[see discussion by ][]{Zhao99}. 
A more exotic possibility suggested by \citet{Maeda02} is 
that the central source was much brighter in X-ray some
300 years ago as a result of the accretion of infalling material
pushed in by the forward shocks from the supernova explosion in Sgr A East,
leading to the ionization of the streamers.

A close physical link between the neutral gas in the CND and the 
ionized gas in Sgr A West is a widely occurring phenomenon, and
this fact can be demonstrated more convincingly in the comparison
of the channel maps as shown in Figure~3.  Within the velocity
ranges of $\pm143$ km s$^{-1}$, the bulk of H41$\alpha$ and
HCO$^+$ features trace the same rotation-like kinematics about 
Sgr A$^\star$.  More importantly, the H41$\alpha$ features frequently 
occur just inside of the HCO$^+$ features 
(see the $V = -$91, $-$39, and +65 km s$^{-1}$ channels).  As 
in the comparison of the velocity integrated images 
(Figure~\ref{fig:hco+}), this physical link is more evident
along the western arc.
There is a tendency for the H41$\alpha$ features to 
show an azimuthal offset from the HCO$^+$ features, 
particularly for the features closer to Sgr A$^\star$
(e.g. V=$-$39 and V=+65 km s$^{-1}$ channels).  
These azimuthal offsets and a broad total line width 
($\pm200$ \kms) for the H41$\alpha$ line are naturally
explained if the gas follows nearly Keplerian orbits around
a highly concentrated mass distribution.
(see the discussion in \S~\ref{sec:model}).

\subsection {Modeling Gas Kinematics in the CND
\label{sec:model}}

Constrained by excellent high spatial resolution imaging 
data in 12.8 $\mu$m [Ne II] and radio recombination lines, 
several plausible and detailed dynamical models for the 
ionized streamers have been constructed 
\citep[e.g. ][]{Sanders98,Vollmer00}. 
Models of gas kinematics for the CND have also been constructed,
mostly to explain the features identified in
low resolution images ($\theta\sim 10-20\arcsec$) that are 
susceptible to confusion with ubiquitous foreground and 
background emission 
\citep[e.g. ][]{Liszt85,Marshall95}.  Taking advantage of
the new high (and matched) resolution spectroscopic imaging
data in H41$\alpha$ and \hcop, we can now examine in detail the 
gas kinematics and the gravitational potential in the central
few parsec radius region in a
self-consistent manner.  We have already concluded that some of
the ionized gas is orbiting \sgrastar\ in the plane defined
by the CND (see \S~\ref{sec:CND}). 
Using the velocity information, we can also isolate and examine 
other kinematic features proposed to be associated with the CND, 
such as the ``+50 \kms\ streamer'' and the ``+70 \kms\ feature'' 
\citep[see ][]{Jackson93}.

If a massive central source 
\citep[2.5-4.1 $\times 10^6$ M$_\odot$, ][]{Genzel99,Ghez00,Ghez03}
dominates the gravitational potential in the central parsec
region, the kinematics of the surrounding gas disk should be 
regular and well behaved.
The neutral and ionized gas inside the CND contributes 
only about 1000 \Mo\ while we estimate the total mass of the
CND to be about $3\times 10^5$ \Mo\ (see \S~\ref{sec:CND}). 
The mass of the central stellar cluster is more substantial and
is highly centrally concentrated, doubling the enclosed total
mass near 2 pc radius \citep{Genzel03}.  We adopt a mass distribution
of the form similar to Eq.~3 of \citet{Vollmer02}, after updating
it using the most recent mass estimates of the massive central
source by \citet[][]{Ghez03} and the dynamical mass estimates for
the central 10 pc radius by \citet[][]{Genzel03}, 
\begin{equation}
M(R) = (4.1 + 1.4 R^{1.25})\times 10^6 M_\odot
\end{equation}
\noindent
where $R$ is radius in pc. The resulting Keplerian-like rotation of 
the gas disk nicely reproduces the observed features in the channel maps
such as the azimuthal displacement for radially distinct 
features as discussed in \S~\ref{sec:CND}.  
To examine whether our interpretations of the gravitational 
potential, the geometry of the CND, and the relationship between
the ionized streamers and the dense neutral gas in the CND are
correct, we have created simulated channel maps of the CND using
a Monte Carlo realization.  They are shown side-by-side in
comparison with the observed channel maps in Figure~11. 

Using Matlab software package, we created a disk of 3000 particles 
with a $r^{-1/2}$ radial distribution.  The model disk has a radial
extent from 0.1 pc to 2.5 pc and has a radially
dependent thickness with an opening angle of $4^\circ$.
This particular disk geometry is adopted
to allow a finite thickness which may play a role in the {\it apparent} 
gas distribution and kinematics in the projected view,
but it is not motivated any particular observational 
constraints.\footnote{The only estimate found in the literature is by
\citet{Latvakoski99} of $\sim7^\circ$ opening angle, based on
their own morphological and dust heating model.}  
The model disk is viewed with 
an inclination angle of 68$^\circ$ and is rotated 
by 60$^\circ$ in the sky plane to match the aspect ratio and the 
position angle of the CND.  Each disk particle is
in rotation about the central potential described by Eq.~5
and has a Gaussian random velocity component of 10 \kms. 
To produce the model channel maps shown in Figure~11,
model disk particles with the line-of-sight velocity matching the 
particular channel velocity window are selected, and their distribution is
convolved with the observing beam ($7.0\arcsec \times 3.5\arcsec$, 
PA$=-5^\circ$).  The only adjustment made to the model was adding
+13 \kms\ to the systemic velocity in order to match the highest
velocity \hcop\ features at $-117$ \kms\ and +143 \kms.

Even though our model disk is constructed for an illustrative purpose
only and is not tuned to match the observations in detail, the 
side-by-side comparisons of the model and data 
shown in Figure~11 offer significant insights on the gas distribution
and kinematics:

\noindent (i) As noted already, some of the ionized gas traced in 
H41$\alpha$ is sharing the same orbital plane as the dense neutral 
gas in the CND (see \S\ref{sec:CND}).  The model disk has 
no knowledge of temperature or ionization information,
but presumably the inner part ($R\lesssim 30''$) of the disk 
is either atomic or ionized.  
The clumps of the H41$\alpha$ emission shown in contours on 
both panels in Figure~11 are frequently overlaying 
the model emission regions shown in grey-scale, especially in 
the channels with velocities between $\pm$ 91 \kms.  The northern
arm is sometimes modeled as being displaced from the plane
defined by the CND \citep[e.g., ][]{Latvakoski99}, but a
large fraction of the northern arm gas seen in the velocity
range between 0 and +143 \kms\ closely follow our kinematical model.

\noindent (ii) The same comparison also shows that the gas in the 
eastern arm and the central bar {\it do not} follow the model. 
This leads us to believe that the ionized gas in the bar and the 
eastern arm is largely concentrated in a different orbital
plane as suggested by others previously \citep[e.g., ][]{Vollmer00}.  

\noindent (iii) The similarity between the observed HCO$^+$ emission 
and the model disk emission is more striking.  
Except for the channels with $-39$ \kms\ $\le$ V $\le +39$ \kms\ 
where foreground absorption has severely altered the appearance 
of the emission features, most of the \hcop\ emission seen 
in the channel maps have matching features in the model.  
The \hcop\ emission arising from
the CND is frequently asymmetric about its major axis, unlike
the model, and the CND is probably not a solid ring or a solid disk
and is likely highly clumped. 
Two channels show substantial \hcop\ emission not
predicted by our model, and these channels correspond to
velocities for the ``+50 \kms\ streamer'' and the ``+70 \kms\ feature''
that are already known to deviate from the CND rotation
(see below).

The +50 \kms\ streamer is a band of emission running
nearly east-west at a slightly oblique angle in the V = +39 \kms\
and V = +65 \kms\ channels, previously noted by \citet{Jackson93}
and others.  This feature extends
well beyond the field of view of the OVRO interferometer and
is several arcmin in length in the
single dish HCN (3--2) channel maps of \citet{Marshall95}.
This is clearly a structure much larger than the CND in size, and thus
the +50 \kms\ streamer is most likely a foreground or a background
structure, not directly associated with the CND.

The +70 \kms\ streamer is the ``7" shaped feature present in 
the V = +91 \kms\ channel.  This structure is also seen in
the single dish HCN (3--2) channel maps of \citet{Marshall95},
but it is more compact than the +50 \kms\ streamer.  Although
this may be a chance projection, this feature lies nearly
exactly over the CND.  However, its velocity differs from the CND 
rotation by about 130 \kms.  It is not as certain that the +70 \kms\ 
streamer is another foreground structure.
Its long, linear appearance is suggestive of a large scale shock 
in origin.  It is not seen in 
H41$\alpha$ or in continuum to offer any further clues, and  
the lack of any ionized gas at the corresponding velocity is
an important argument against the +70 \kms\ streamer being part of the CND.

One of the key arguments against the 
CND being a coherent disk or a ring is the apparent gap 
along its eastern side in the previous HCN and HCO+ maps.  This gap
now appears filled with several compact gas clumps such as the clump I 
in our new HCO+ image shown in Figure~\ref{fig:hco+}.  
In addition, neutral gas with higher temperature and density 
also fills this gap as inferred by the observations of  
[C II] and NH$_3$ \citep[see ][]{Jackson93,Herrnstein02}.  
Therefore earlier arguments based on the apparent morphology 
in the lower sensitivity HCN and HCO+ maps are no
longer valid.  The mean velocity of the warm
molecular gas found by Herrnstein \& Ho (their clump D)
is slightly redshifted ($<V>\sim +50$ \kms) than expected from our 
simple rotation model, but this velocity shift is in the correct sense
if their average spectrum includes the emission from the entire
north-south extent of their clump D.  

In summary, we have successfully demonstrated the 
continuity in the distribution and kinematics of the ionized and the
neutral gas within the central 3 pc diameter region surrounding 
\sgrastar\ using our new observational data and a simple kinematical model. 
Our kinematical model reproduces 
many of the observed neutral and ionized gas features, as shown
in Figure~11.  The success of this simple model strengthens the 
notion that most of the gas found in the central parsec of \sgrastar\
partake in a simple rotation around the central source
within the disk plane defined by the CND.
Our kinematical model also identifies very clearly which
gas components {\it do not} follow this simple rotation.
In addition, our model suggests that
some of the subtle observed effects such as the slight
asymmetry between the near and far side of the disk are the
products of the finite size, asymmetric observing beam and 
projected viewing of the 3-D CND structure (e.g., see $V=\pm39$ \kms\
channel).  
The overall picture of the gas distribution and kinematics in
the central 3 pc surrounding \sgrastar\ is not yet complete because
information on the distribution and kinematics of 
atomic gas inside the CND is quite incomplete.  Future arcsecond
resolution imaging of dense atomic gas tracers such as [C II]
will provide important clues for completing our understanding
of the gaseous environment in this region.

\section{SUMMARY}

The OVRO millimeter array observations of the H41$\alpha$ and  
92 GHz continuum emission from the ionized streamers around \sgrastar\ 
(Sgr A West) and the HCO$^+$ (1--0) emission from the 
circumnuclear disk (CND) are presented along with a new survey
of the hydrogen recombination lines at millimeter wavelengths
(H31$\alpha$, H41$\alpha$, and H44$\alpha$) obtained using the 
NRAO 12-m telescope.  These observations are 
compared with the VLA H92$\alpha$ and the 8.3 GHz continuum data 
by Roberts \& Goss (1993).   We summarize our results as following:

\noindent 1. The 92 GHz continuum image is essentially identical 
to the VLA continuum image at 8.3 GHz in morphology.  The total
continuum flux detected inside the 85\arcsec\ diameter is about
12 Jy, but some extended flux associated with structure larger 
than about 30\arcsec\ is missed.  The brightest feature, \sgrastar,
is unresolved at 3\arcsec\ resolution of our observations and has 
a peak flux density of $1.95\pm0.30$ Jy beam$^{-1}$.  
Assuming the 7000 K electron gas fills our 
0.005 pc (1000 AU) beam, we infer average continuum 
optical depth of $\tau \lesssim 10^{-3}$ and electron density 
$n_e \sim 2\times 10^4$ cm$^{-3}$.

\noindent 2. The kinematics and the morphology of 
H41$\alpha$ emission are very similar to 
those observed in H92$\alpha$ and 12.8 $\mu$m [Ne~II] lines.  The brightest 
emission occurs along the north arm with a peak line brightness 
temperature of less than 10 K.  This suggests a H41$\alpha$ optical 
depth of $\tau_L \lesssim 10^{-3}$ averaged over the 13 \kms\ channel width.
The emission along the central bar displays a line-of-sight velocity 
ranging over 300 \kms, and this gas may be located in the deepest 
part of the gravitational potential within the CND region.

\noindent 3. The molecular CND is imaged in HCO$^+$  
with a 40\% improvement in resolution and 4-15 times better
sensitivity over the previous BIMA observations made in HCN and \hcop. The
CND consists of several discrete concentrations, and their
peak brightness ranging between 10 and 30 K averaged over 
the 0.1 pc synthesized beam suggests that they originate from 
the warm (60-100 K) gas responsible for NH$_3$ and dust 
continuum emission in the CND.  A virial analysis suggests 
that these clouds have mean density of $n\sim 10^7$ cm$^{-3}$,
sufficient to withstand the tidal shear in the region, and the
CND may be a long lived structure.  We estimate a total gas mass of 
about $3\times 10^5 M_\odot$ for the CND.  Our non-detection of
\hcop\ absorption along the \sgra\ West places an upper limit
to the gas column density of $\lesssim 4 \times 10^{21}$ cm$^{-2}$,
or the neutral gas within the CND is depleted of molecules. 

\noindent 4. A line-to-continuum ratio analysis under the 
LTE assumption yields a fairly uniform electron temperature 
of about 7000 K along the ionized streamers in the Sgr A West 
complex, in an excellent agreement with the electron temperature
calculated at 8.3 GHz using the H92$\alpha$ observations \citep{Roberts93}.
Evidence for increased $T_e$ (or an unusually low L/C ratio) 
along the central bar just southwest of \sgrastar\ is 
found in our data as well even though our data include all line 
emission within $\pm 350$ \kms.  The low L/C ratio we find at 92 GHz 
challenges the optically thick continuum sources IRS~2 and IRS~13 
as additional sources of continuum emission as suggested by \citet{Roberts96}. 
Therefore, the elevated $T_e$ (a lower L/C ratio) seen near the 
location of IRS~2 and IRS~13 is still not fully understood.

\noindent 5. A comparison of the velocity integrated line flux 
densities for H31$\alpha$, H41$\alpha$, H44$\alpha$, and H92$\alpha$
emission from the \sgra\ West complex shows line ratios characteristic 
of spontaneous emission from optically thin gas near LTE.  
A detailed analysis reveals that observed recombination line brightness 
is reduced by 10-30\% (i.e., $b_n\sim 0.7-0.9$) from the LTE value.

\noindent 6. A combined analysis of the new high resolution 
H41$\alpha$ and \hcop\ images and a kinematic model clearly
demonstrates a widely spread, close physical and dynamical 
link between the molecular gas in the CND and the ionized gas,
including along the eastern rim where a gap in the CND was
previously reported.  The same analysis also clearly
identifies several features that do not follow the disk 
rotation.  We conclude that the ``+50 \kms\ streamer'' is
not directly associated with the CND.  The morphology of
``+70 \kms\ streamer'' is suggestive of its association 
with the CND, but its kinematics and absence of any
ionized gas challenge the idea of their close association.

\acknowledgements

We are grateful to D. Roberts 
for providing us with the H92$\alpha$ and 8.4 GHz continuum data 
from the VLA for comparison.  We are indebted to C. Lang for her expert
advice and help with the GIPSY data reduction package and for
many helpful suggestions as the referee of this article. 
In addition, we would like to acknowledge
the direction gained from insightful discussions with W. M. Goss and 
late K. R. Anantharamaiah.  We would like to thank the OVRO and the NRAO
12-m Telescope staff for the support and help without which this 
project would not have been possible.
The National Radio Astronomy Observatory is a facility of the 
National Science Foundation operated under cooperative agreement 
by Associated Universities, Inc.  

\bigskip
\bigskip


\begin{figure}
\figurenum{3}
\caption{(a) Channel maps (from $-$221 \kms\ to $-$13 \kms) of HCO$^+$ 
emission are shown in grey-scale and H41$\alpha$ in contours.  
(b) Channel maps (from +13 \kms\ to +221 \kms) of HCO$^+$ 
emission are shown in grey-scale and H41$\alpha$ in contours.
The grey-scale ranges between $-$100 (white) and +500 (black) mJy 
beam$^{-1}$ for HCO$^+$.  The contour levels are $-$2, +2, +3, +4, +5, 
and +6 times 15 mJy beam$^{-1}$ ($1\sigma$).  The location of 
\sgrastar\ is marked at the center with a cross.
\label{fig:ch}}
\end{figure}


\begin{figure}
\figurenum{4}
\plotone{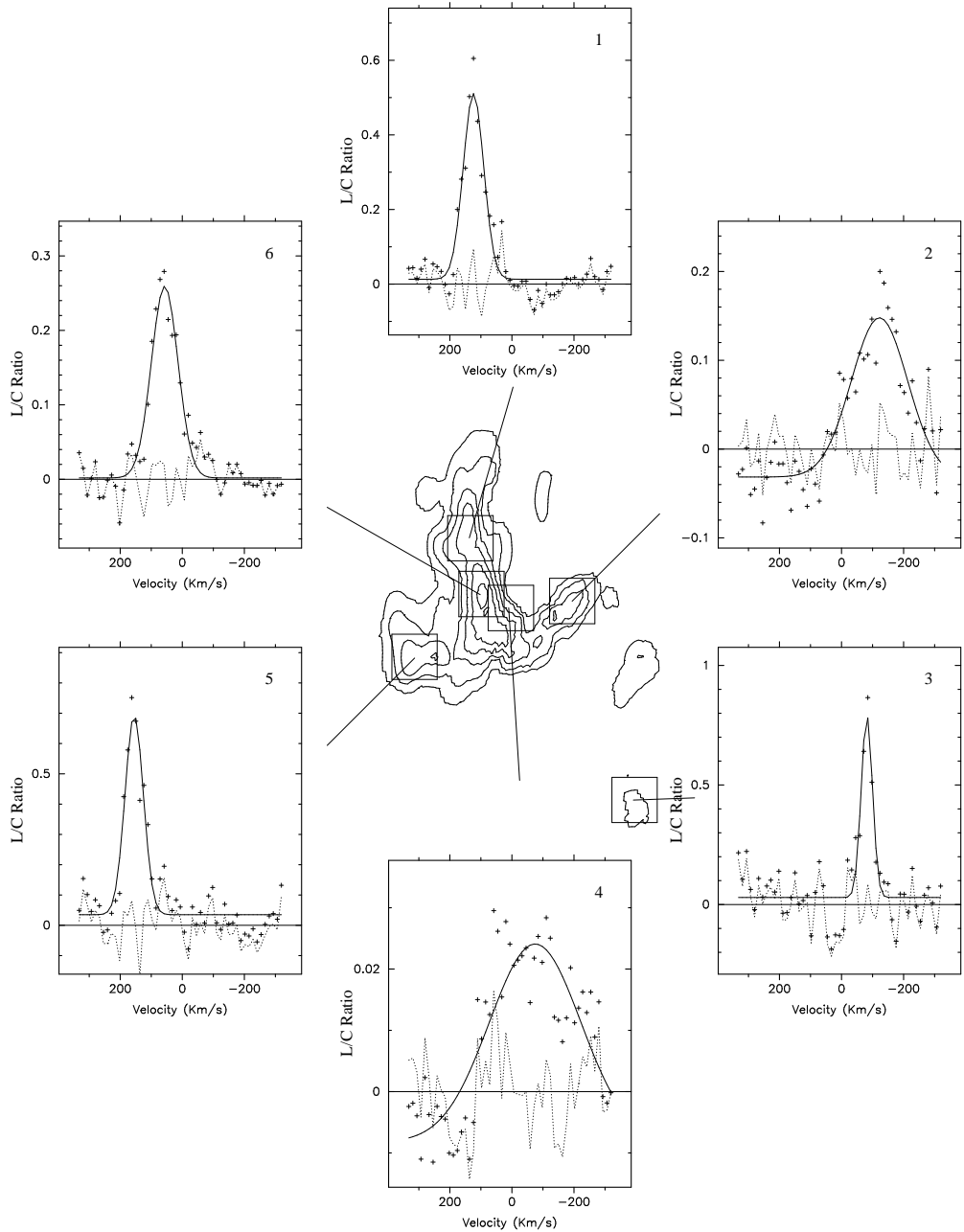}
\caption{A contour map of H41$\alpha$ emission is shown with line-to-continuum 
(L/C) plots of respective regions. The contour levels are +1, +2, +3, +4, 
+5, and +6 times 2.3 Jy \kms. Ratios of the L/C are marked on the 
y-axis and the velocity ranges in \kms\ are shown on the x-axis. 
Crosses mark the data points, solid line
represents the model Gaussian fit, and the dashed line represents the residual.
\label{fig:L/C}}
\end{figure}

\begin{figure}
\figurenum{5}
\plotone{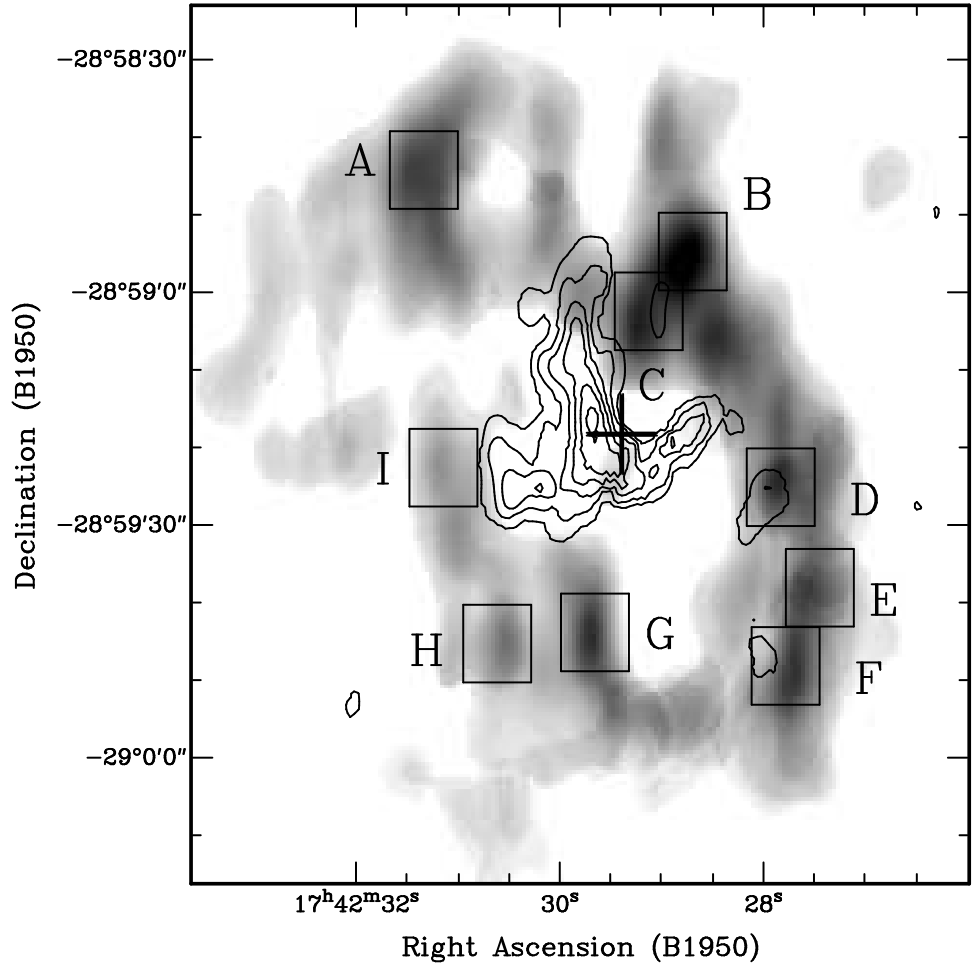}
\caption{Integrated line emission of the HCO$^+$ line from \sgra\ complex at 
$6.95\arcsec \times 3.47\arcsec$ (PA = $-5^\circ$) resolution. 
The cross marks the position of \sgrastar. The grey-scale range is 
between 7 to 32 Jy km s$^{-1}$ beam$^{-1}$. 
Superimposed are the contours of H 41$\alpha$ emission. The contour 
levels are 2.3, 4.6, 6.9, 9.2, 11.6, and 13.9 Jy km s$^{-1}$ beam$^{-1}$.
The boxes mark the dense gas clumps described in Table~\ref{tab:hco+}. 
\label{fig:hco+}}
\end{figure}

\begin{figure}
\figurenum{6}
\plotone{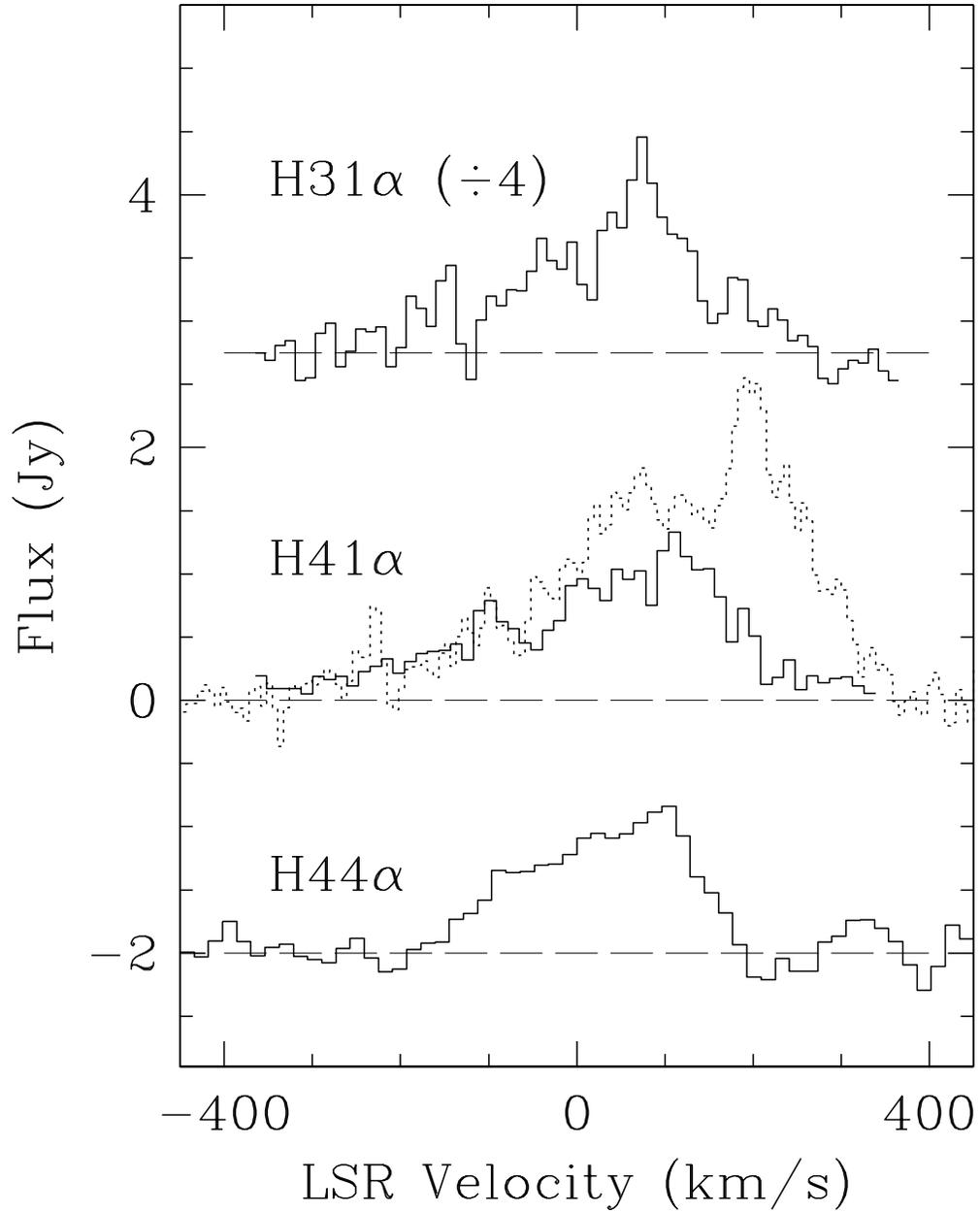}
\caption {H31$\alpha$, H41$\alpha$, and H44$\alpha$ spectra of
Sgr A West obtained with the NRAO 12-m telescope.  The H31$\alpha$ spectrum
is made by summing 8 independent spectra covering the whole Sgr A West,
scaled down by a factor of four.
Both the OVRO (solid line) and the NRAO 12-m (dotted line) spectrum
are shown for H41$\alpha$ line.  The excess emission in the NRAO 12-m
spectrum near +200 \kms\ may be due the contamination by widely
distributed foreground CH$_3$CN emission (see text).
\label{fig:12m}}
\end{figure}

\begin{figure}
\figurenum{7}
\plotone{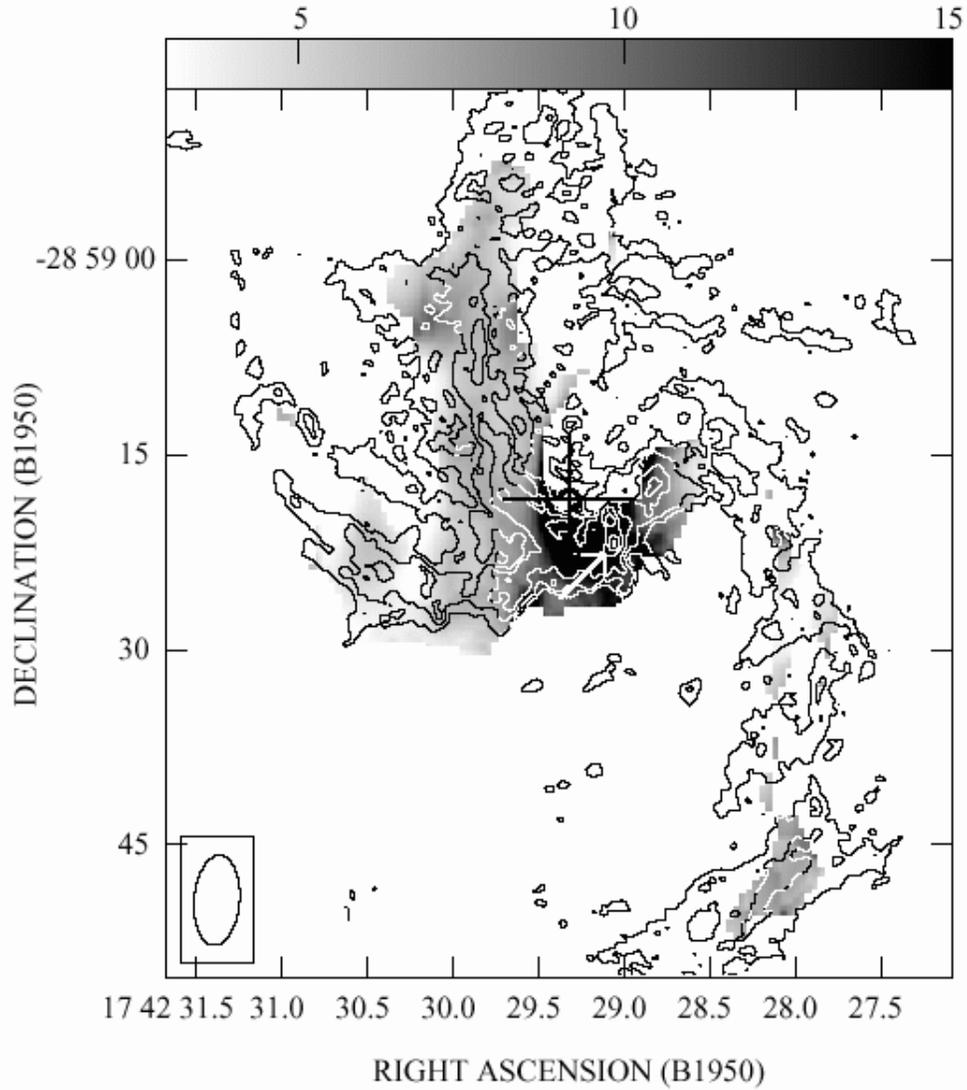}
\caption{Grey-scale representation of LTE electron temperature 
distribution in Sgr A West.  The grey-scale ranges between 3000 to 15000 K.
The 8.3 GHz continuum image at $0.62''\times 0.51''$ resolution
by \citet{Roberts93} is shown in contours.  The contour levels are
2.5, 5, 10, 20, 30, 40, 50, and 60 mJy beam$^{-1}$.
The thick cross marks the position of \sgrastar.  The location of
the IR source IRS~2 is shown with a thick white arrow, and the
second continuum source about $2''$ to the north is IRS~13.
\label{fig:Te}}
\end{figure}

\begin{figure}
\figurenum{8}
\plotone{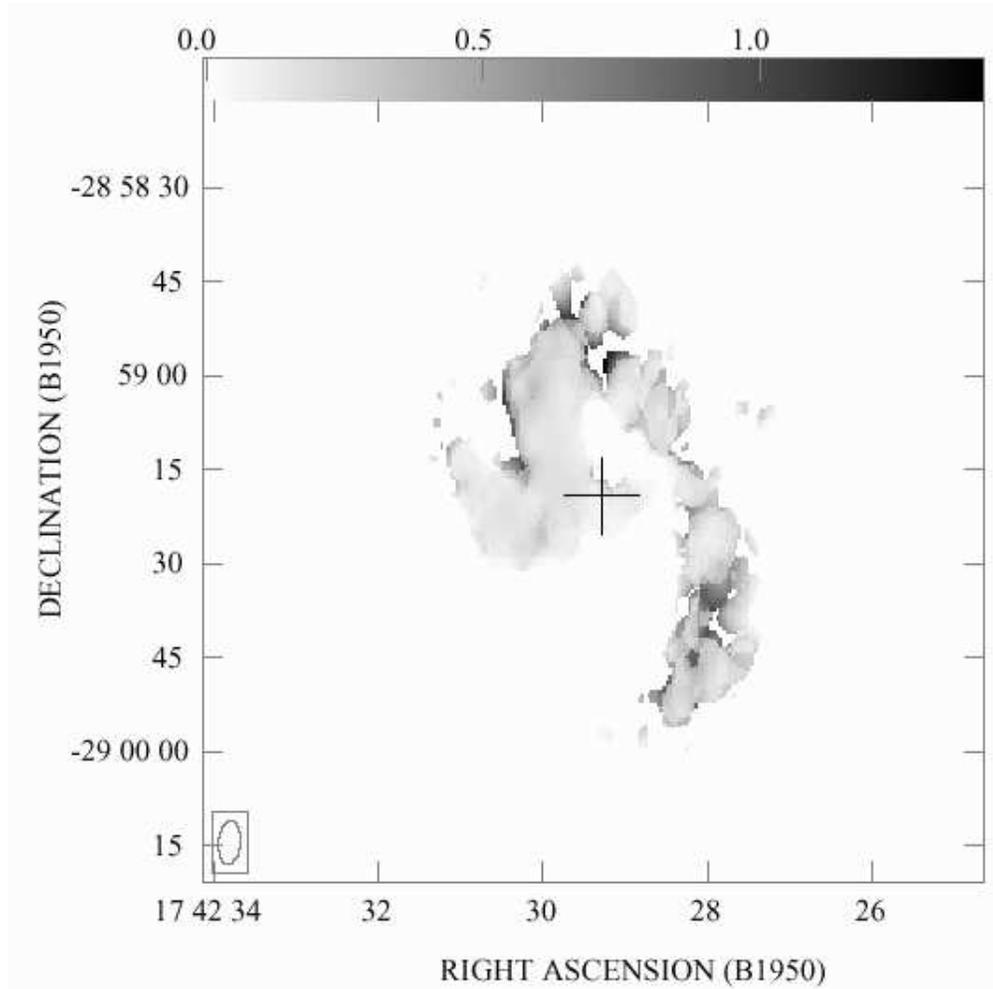}
\caption {A ratio map between H92$\alpha$ and H41$\alpha$ line emission 
is shown in grey-scale (from 0.0 to 1.5). The 
integrated H92$\alpha$ line data from Roberts et al. (1993) were
convolved with the OVRO beam.  The ratio is nearly constant,
$\sim$0.15, in most regions. 
\label{fig:H92a/H41a}}
\end{figure}

\begin{figure}
\figurenum{9}
\plotone{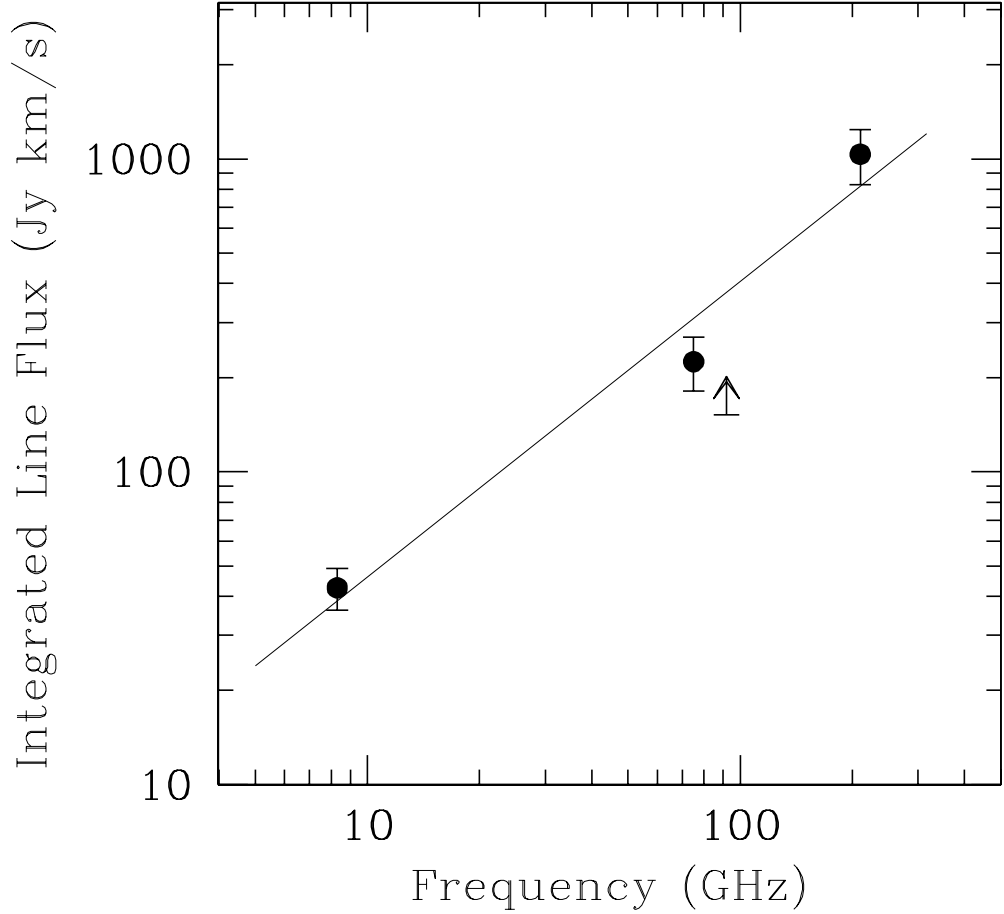}
\caption{Velocity integrated line flux densities of
H31$\alpha$, H44$\alpha$, and H92$\alpha$ are plotted   
as a function of frequency.  The H41$\alpha$ measurement from
the OVRO data is shown only as an upper limit since some
of the extended emission is resolved out by the interferometer. 
The solid line is the best fit power-law relation, $S_L\propto
\nu^{\alpha}$.  The slope $\alpha=0.95\pm0.18$ is very close
to the theoretical value for spontaneous emission 
($S_L\propto \nu^{+1}$).
\label{fig:LTE}}
\end{figure}


\begin{figure}
\figurenum{10}
\includegraphics[angle=270, scale=0.8]{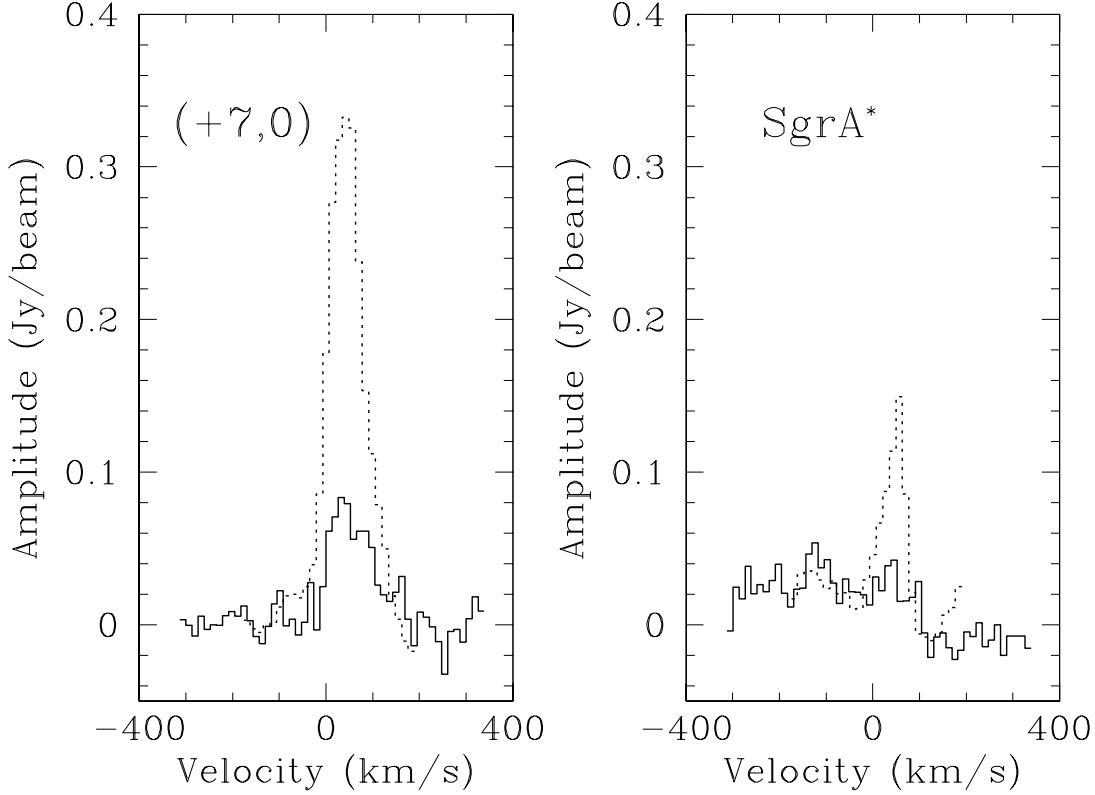} 
\caption {A comparison of the H41$\alpha$ (solid line)
and H92$\alpha$ (dotted line, multiplied by a factor of 15)
spectrum, one at the Sgr A$^\star$ position (right) and at an
adjacent position 7\arcsec\ to the east (left).
\label{fig:twospecs}}
\end{figure}

\begin{figure}
\figurenum{11}
\caption{(a) Channel maps (from $-$195 \kms\ to $-$13 \kms) of HCO$^+$ 
emission are shown in grey-scale.  In each twin panel plot, the
observed HCO$^+$ emission is shown on the left panel while the
simulated disk model (see \S~\ref{sec:model}) is 
shown on the right panel.  The contours of H41$\alpha$ emission are
shown in both panels.   The grey-scale
ranges between $-$100 (white) and +500 (black) mJy beam$^{-1}$ for
HCO$^+$.  The greyscale for the model shown on the right panel 
uses the full range in model units.
The contour levels are $-$2, +2, +3, +4, +5, and +6 times 15 mJy
beam$^{-1}$ ($1\sigma$).  Each box is 100\arcsec\ on each side,
and \sgrastar\ is located at the center.
(b) Channel maps (from +13 \kms\ to +195 \kms) of HCO$^+$ in 
grey-scale superimposed with contours of H41$\alpha$.
\label{fig:model}}
\end{figure}



\begin{deluxetable}{lc}
\tablewidth{0pt}
\tablecaption{Summary of the OVRO Observations \label{tab:ovrosummary}}
\tablehead{}
\startdata
Rest Frequency of H41$\alpha$ line & 92.034 GHz \\
Rest Frequency of HCO$^+ (J = 1 \rightarrow 0)$ line & 89.188 GHz \\
Primary beam (HPBW)		    	    &	 85$\arcsec$	 \\
Phase center 	    &	$\alpha(B1950)=17^h 51^m 29.4^s$ \\
                    &   $\delta(B1950)= -28^\circ 59' 19.2''$ \\
Center LSR velocity &	0.0 \kms \\
Synthesized beam (HPBW)		    &	$6.95\arcsec \times 3.47\arcsec$ (PA$=-5^\circ$)\\
Spectrometer settings 		    &	$60 \times 4$ MHz (13 \kms) \\
RMS noise 			    & 0.13 Jy beam$^{-1}$ (0.21 K)\\
\enddata
\end{deluxetable}


\begin{deluxetable}{llllcccc}
\tablewidth{0pt}
\tablecaption{92 GHz continuum emission \label{tab:continuum}}
\tablehead{
\multicolumn{1}{c}{Feature} &
\multicolumn{1}{c}{Size} &
\multicolumn{1}{c}{$S_{peak}$} &
\multicolumn{1}{c}{$T_{peak}$} &
\multicolumn{1}{c}{$S_{total}$} \\
\multicolumn{1}{c}{} &
\multicolumn{1}{c}{(pc)} &
\multicolumn{1}{c}{(Jy Beam$^{-1}$)} &
\multicolumn{1}{c}{(K)} &
\multicolumn{1}{c}{(Jy)} \\}
\startdata
Northern Arm 	& 0.9 & $0.39 \pm 0.07$ & $3.1\pm0.6$  & $1.09 \pm 0.16$ \\
\sgrastar\	& 0.1 & $1.95 \pm 0.29$ & $15.6\pm2.3$ & $1.95 \pm 0.30$ \\ 
Bar	     	& 1.1 & $0.75 \pm 0.14$ & $6.0\pm1.1$  & $0.90 \pm 0.29$  \\
Western Arc   	& 1.5 & $0.14 \pm 0.02$ & $1.1\pm0.2$  & $1.08 \pm 0.08$  \\
Eastern Arm 	& 0.4 & $0.29 \pm 0.05$ & $2.3\pm0.4$  & $0.52 \pm 0.12$  \\ 
\enddata
\end{deluxetable}


\begin{deluxetable}{llllcccc}
\tablewidth{0pt}
\tablecaption{A Summary of H41$\alpha$ line intensity \label{tab:h41a}}
\tablehead{
\multicolumn{1}{c}{Region} &
\multicolumn{1}{c}{Size} &
\multicolumn{1}{c}{$S_{peak}$} &
\multicolumn{1}{c}{$T_{peak}$} &
\multicolumn{1}{c}{$S_{total}$} \\
\multicolumn{1}{c}{} &
\multicolumn{1}{c}{(pc)} &
\multicolumn{1}{c}{(Jy Beam$^{-1}$)} &
\multicolumn{1}{c}{(K)} &
\multicolumn{1}{c}{(Jy)} \\}
\startdata
Northern Arm 	& 0.9 & $1.18 \pm 0.27$ & $9.4\pm2.2$ & $4.70 \pm 0.71$  \\
Bar	     	& 1.1 & $1.12 \pm 0.21$ & $9.0\pm1.7$ & $5.70 \pm 0.37$  \\
Western Arc   	& 1.5 & $0.50 \pm 0.14$ & $4.0\pm1.1$ & $1.86 \pm 0.39$  \\
Eastern Arm	& 0.4 & $0.90 \pm 0.15$ & $7.2\pm1.2$ & $2.02 \pm 0.36$  \\
\enddata
\end{deluxetable}

\clearpage

\begin{deluxetable}{ccccccc}
\tablewidth{0pt}
\tablecaption{LTE electron temperature of the regions identified in 
Figure 4 \label{tab:Te}}
\tablehead{
\multicolumn{1}{c}{Region} &
\multicolumn{1}{c}{Box Center} &
\multicolumn{1}{c}{$V_{LSR}$} &
\multicolumn{1}{c}{$\Delta V$} &
\multicolumn{1}{c}{$<T{_L}$/$T{_C}>$} &
\multicolumn{1}{c}{Temperature $T^{\star}_e$}\\
\multicolumn{1}{c}{} &
\multicolumn{1}{c}{(J2000)} &
\multicolumn{1}{c}{(km s$^{-1}$)} &
\multicolumn{1}{c}{(km s$^{-1}$)} &
\multicolumn{1}{c}{} &
\multicolumn{1}{c}{(K)}\\
}
\startdata
1 & $17^h42^m29.^s71,~~-28^\circ59'09.''2$ & $+114 \pm 3$  & 78  $\pm$ 5  & 0.50 $\pm$ 0.04 & 
6394$\pm$608 \\
2 & $17^h42^m28.^s60,~~-28^\circ59'18.''2$ & $-132 \pm 8$  & 213 $\pm$ 18 & 0.18 $\pm$ 0.02 & 
6473$\pm$891 \\
3 & $17^h42^m27.^s92,~~-28^\circ59'46.''7$ & $-92 \pm  3$  & 43  $\pm$ 5  & 0.80 $\pm$ 0.13 & 
7470$\pm$1372 \\
4 & $17^h42^m29.^s27,~~-28^\circ59'19.''2$ & $-117 \pm 20$ & 347 $\pm$ 48 & 0.03 $\pm$ 0.01 & 
18806$\pm$4770 \\
5 & $17^h42^m30.^s32,~~-28^\circ59'26.''2$ & $+142 \pm 3$  & 70  $\pm$ 5  & 0.66 $\pm$ 0.07 & 
5528$\pm$623 \\
6 & $17^h42^m29.^s59,~~-28^\circ59'17.''2$ & $+46 \pm 3$   & 101 $\pm$ 6  & 0.26 $\pm$ 0.02 & 
9057$\pm$825 \\
\enddata
\end{deluxetable}


\begin{deluxetable}{cccccccc}
\tablewidth{400pt}
\tablecaption{HCO$^+$ line intensity of the nine brightest clumps in the CND
\label{tab:hco+}}
\tablehead{
\multicolumn{1}{c}{Clump} &
\multicolumn{1}{c}{Size} &
\multicolumn{1}{c}{$S_{peak}$} &
\multicolumn{1}{c}{$T_{peak}$} &
\multicolumn{1}{c}{$\Delta V$} &
\multicolumn{1}{c}{$M_{vir}$} &
\multicolumn{1}{c}{$\bar{n}_{vir}$} \\
\multicolumn{1}{c}{ } &
\multicolumn{1}{c}{(pc)} &
\multicolumn{1}{c}{(Jy Beam$^{-1}$)} &
\multicolumn{1}{c}{(K)} &
\multicolumn{1}{c}{(\kms)} &
\multicolumn{1}{c}{(M$_\odot$)} &
\multicolumn{1}{c}{(cm$^{-3}$)} \\}
\startdata
A & 0.4 & $3.5 \pm 0.8$ & $27.9\pm6.4$ & 20.7 & 2.1  $\times 10^4$ & $1.4\times 10^7$  \\
B & 0.3 & $1.6 \pm 0.4$ & $12.7\pm3.2$ & 17.3 & 1.1  $\times 10^4$ & $1.8\times 10^7$  \\ 
C & 0.3 & $1.0 \pm 0.8$ & $8.0\pm6.4$  & 12.0 & 0.5  $\times 10^4$ & $0.9\times 10^7$ \\
D & 0.4 & $1.2 \pm 0.3$ & $9.5\pm2.3$  & 29.9 & 4.5  $\times 10^4$ & $3.0\times 10^7$ \\
E & 0.2 & $1.5 \pm 0.9$ & $11.9\pm7.2$ & 11.6 & 0.3  $\times 10^4$ & $1.8\times 10^7$  \\
F & 0.3 & $1.9 \pm 0.4$ & $15.1\pm3.2$ & 19.4 & 1.4  $\times 10^4$ & $2.2\times 10^7$  \\ 
G & 0.6 & $1.5 \pm 0.4$ & $12.0\pm3.2$ & 16.1 & 1.9  $\times 10^4$ & $0.4\times 10^7$ \\
H & 0.4 & $1.0 \pm 0.5$ & $8.0\pm4.0$  & 19.0 & 1.8  $\times 10^4$ & $1.2\times 10^7$  \\
I & 0.2 & $1.8 \pm 0.7$ & $14.3\pm5.6$ & 13.0 & 0.4  $\times 10^4$ & $2.3\times 10^7$  \\ 
\enddata
\end{deluxetable}


\begin{deluxetable}{lcc}
\tablewidth{0pt}
\tablecaption{Summary of the NRAO 12-m Observations \label{tab:12m}}
\tablehead{
\multicolumn{1}{l}{Line} &
\multicolumn{1}{c}{$\nu$} &
\multicolumn{1}{c}{$S \Delta V$}\\
\multicolumn{1}{l}{} &
\multicolumn{1}{c}{(GHz)} &
\multicolumn{1}{c}{(Jy \kms)}\\}
\startdata
H31$\alpha$ &  210.502  &  $1034\pm207$ \\
H35$\alpha$ &  146.969  & -- \\
H41$\alpha$ &  92.034  & -- \\
H44$\alpha$ &  74.645  &  $232\pm46$ \\
H92$\alpha$ &   8.309  &  $42.6\pm6.5$ \\
\enddata
\end{deluxetable}

\end{document}